\documentclass[traditabstract]{aa} % for the abstract without structuration (traditional abstract) 

\usepackage[fleqn]{amsmath}
\usepackage{amssymb}
\usepackage{textcomp}
\usepackage{graphicx}
\usepackage{txfonts}
\usepackage{natbib}
\usepackage{url}
\usepackage{multirow}
\usepackage{subfigure}
\usepackage[usenames,dvipsnames]{xcolor}

\newcommand{\pop}[1]{\langle {\rm #1} \rangle}
\newcommand{\num}[1]{n({\rm #1})}
\newcommand{\ab}[1]{x({\rm #1})}
\newcommand{\op}[1]{{\rm OPR(#1)}}
\newcommand{\opini}[1]{{\rm OPR_0(#1)}}
\newcommand{\opst}[1]{{\rm OPR_{st}(#1)}}

\newcommand{\phh}{{\rm p\mathchar`- H_{2}}}
\newcommand{\ohh}{{\rm o\mathchar`- H_{2}}}
\newcommand{\phhdp}{{\rm p\mathchar`- H_{2}D^{+}}}
\newcommand{\ohhdp}{{\rm o\mathchar`- H_{2}D^{+}}}
\newcommand{\phhhp}{{\rm p\mathchar`- H_{3}^{+}}}
\newcommand{\ohhhp}{{\rm o\mathchar`- H_{3}^{+}}}
\newcommand{\ddt}{\frac{d}{dt}}
\newcommand{\bb}[1]{\left( #1 \right)}
\newcommand{\react}[4]{{\rm #1} + {\rm #2} \rightarrow {\rm #3} + {\rm #4}}
\newcommand{\reactalign}[4]{{\rm #1} + {\rm #2} &\rightarrow {\rm #3} + {\rm #4}}
\newcommand{\reacteq}[4]{{\rm #1} + {\rm #2} \rightleftharpoons {\rm #3} + {\rm #4}} 

\newcommand{\hdo}{{\rm HDO/H_{2}O}}

\newcommand{\phdo}{p_{\rm OH\rightarrow HDO}}

\begin{document}

\title{Water deuteration and ortho-to-para nuclear spin ratio of H$_2$ in molecular clouds formed via the accumulation of \mbox{\ion{H}{i}} gas}
\titlerunning{Water deuteration and ortho-to-para ratio of H$_2$ in molecular clouds}
\authorrunning{Furuya et al.}

\author{K. Furuya\inst{\ref{inst1}} 
\and Y. Aikawa\inst{\ref{inst2}} 
\and U. Hincelin\inst{\ref{inst3}} 
\and G. E. Hassel\inst{\ref{inst4}}
\and E. A. Bergin\inst{\ref{inst5}} 
\and A. I. Vasyunin\inst{\ref{inst6},\ref{inst7}}
\and Eric Herbst\inst{\ref{inst3},\ref{inst8}}}

\institute{Leiden Observatory, Leiden University, P.O. Box 9513, 2300 RA, The Netherlands\\
\email{furuya@strw.leidenuniv.nl}\label{inst1} 
\and Center for Computer Sciences, University of Tsukuba, Tsukuba 305-8577, Japan\label{inst2}
\and Department of Chemistry, University of Virginia, Charlottesville, VA 22904, USA\label{inst3}
\and Department of Physics \& Astronomy, Siena College, Loudonville, NY, 12211, USA\label{inst4}
\and Department of Astronomy, University of Michigan, 500, Church St, Ann Arbor, MI 48109\label{inst5}
\and Max Planck Institut f\"ur Extraterrestrische Physik, Giessenbachstrasse 1, 85748 Garching, Germany\label{inst6}
\and Ural Federal University, Ekaterinburg, Russia\label{inst7}
\and Department of Astronomy, University of Virginia, Charlottesville, VA 22904, USA\label{inst8}}

%\date{}

% \abstract{}{}{}{}{} 
% 5 {} token are mandatory
 
\abstract
{We investigate the water deuteration ratio and ortho-to-para nuclear spin ratio of H$_2$ ($\op{H_2}$) during the formation and 
early evolution of a molecular cloud, 
following the scenario that accretion flows sweep and accumulate \mbox{\ion{H}{i}} gas to form molecular clouds.
We follow the physical evolution of post-shock materials using a one-dimensional shock model, 
combined with post-processing gas-ice chemistry simulations.
This approach allows us to study the evolution of the $\op{H_2}$ and water deuteration ratio without an arbitrary assumption 
of the initial molecular abundances, including the initial $\op{H_2}$.
When the conversion of hydrogen into H$_2$ is almost complete the $\op{H_2}$ is already much smaller than the statistical value of three
because of the spin conversion in the gas phase.
As the gas accumulates, the $\op{H_2}$ decreases in a non-equilibrium manner.
We find that water ice can be deuterium-poor at the end of its main formation stage in the cloud, 
compared to water vapor observed in the vicinity of low-mass protostars where water ice is sublimated.
If this is the case, the enrichment of deuterium in water should mostly occur at somewhat later evolutionary stages of star formation, 
i.e., cold prestellar/protostellar cores.
The main mechanism to suppress water ice deuteration in the cloud is the cycle of photodissociation and reformation of water ice, 
which efficiently removes deuterium from water ice chemistry.
The removal efficiency depends on the main formation pathway of water ice.
The $\op{H_2}$ plays a minor role in water ice deuteration at the main formation stage of water ice.}


\maketitle

\section{Introduction}
\label{sec:intro}
It is well established that water is predominantly formed via grain surface reactions, and is already abundant 
(10$^{-4}$ with respect to hydrogen nuclei) in molecular clouds with line of sight visual extinction $\gtrsim 3$ mag \citep{whittet93,whittet03}.
One of the major unresolved questions in the field of astrochemistry is whether, and how much, water in the solar system originated from 
the parent molecular cloud \citep[e.g., recent reviews by][]{ceccarelli14,vandishoeck14}.
The $\hdo$ ratio can be a useful tool to probe the processing of water during star- and planet-formation.
We can trace water evolution by observing and comparing the $\hdo$ ratio in objects at different evolutionary stages.
The upper limits of the $\hdo$ ice ratio in embedded protostars have been derived as $(2-10)\times10^{-3}$ 
depending on the source \citep{dartois03,parise03},
which are much larger than the elemental abundance of deuterium ($1.5\times10^{-5}$) in the local interstellar medium \citep[ISM,][]{linsky03}.
Recent interferometric observations have revealed the deuteration ratios of water vapor in the inner hot regions ($T>100$ K) of the embedded protostellar sources, 
where water ice is sublimated \citep[][]{jorgensen10,coutens13,coutens14a,persson13,persson14,taquet13a}.
It has been found that the $\hdo$ ratio is as high as 10$^{-3}$, which is similar to that in Oort cloud comets \citep[e.g.,][]{villanueva09,bockelee-morvan12}. 
This similarity may imply that some of the cometary water originated from the embedded protostellar phase or earlier, 
while some processing may have been at work in the disk, taking the variation of the $\hdo$ ratio in the solar system objects into consideration 
\citep[see e.g., recent theoretical work by][]{willacy09,visser11,yang13,furuya13,albertsson14a,cleeves14}.

\begin{figure}
\resizebox{\hsize}{!}{\includegraphics{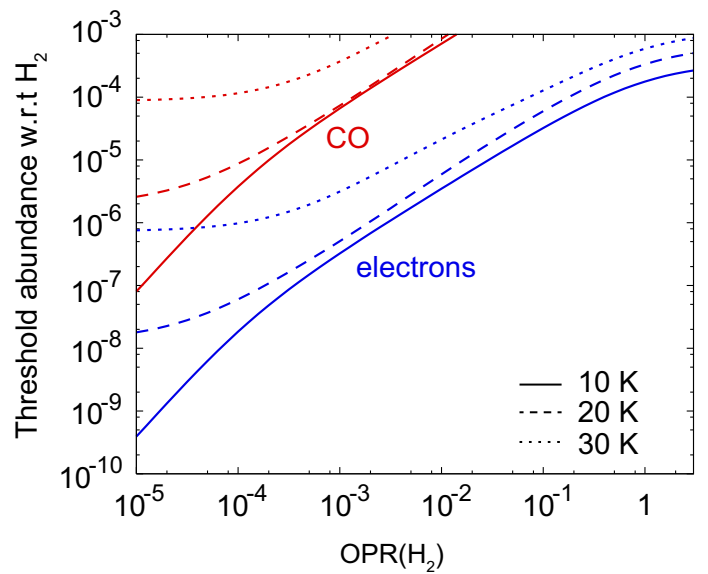}}
\caption{Threshold abundances of CO and electrons to be the dominant reactant of H$_2$D$^+$. If CO and electron abundances with 
respect to H$_2$ are below the lines, Reaction (\ref{eq:dfrac_reaction}) in the backward direction dominates over 
H$_2$D$^+$ destruction by CO and electrons. See Appendix \ref{appendix:opr_dh} for more details.}
\label{fig:critical_opr}
\end{figure}

On the other hand, to the best of our knowledge, there has been no detection of HDO ice in molecular clouds and prestellar cores.
This limits our understanding of water evolution from molecular clouds to the embedded sources.
Instead, there have been numerous theoretical studies on the $\hdo$ ice ratio 
in cold ($\sim$10 K) and dense ($>$10$^4$ cm$^{-3}$) conditions \citep[e.g., recent work by][]{taquet13b,sipila13,lee15}.
When the primary reservoir of deuterium is HD, deuterium fractionation is driven by isotope exchange reactions, such as 
\begin{align}
\reacteq{H_3^+}{HD}{H_2D^+}{H_2}.\label{eq:dfrac_reaction}
\end{align}
The reaction in the backward direction is endothermic, and thus prohibited at low temperatures of $\lesssim$20 K.
The enrichment of deuterium in e.g., H$_3^+$ is transfered to both gaseous and icy molecules through sequential chemical reactions \citep{millar89,roberts04}.
However, the presence of ortho-H$_2$ ($\ohh$) can change the situation.
It can suppress the overall deuteration processes even at $T \lesssim 20$ K, 
since the internal energy of $\ohh$, which is 170.5 K above that of para-H$_2$ ($\phh$), helps to overcome the endothermicity of 
the exchange reactions in the backward direction \citep{pagani92,gerlich02,walmsley04}.
CO and electrons also play a role in deuterium chemistry; they destroy H$_2$D$^+$ to suppress further fractionation \citep{roberts02}.
Figure \ref{fig:critical_opr} shows the threshold abundances of CO and electrons with respect to H$_2$ as functions of 
the ortho-to-para nuclear spin ratio of H$_2$ ($\op{H_2}$);
below the lines for CO and electrons in this figure, 
Reaction (\ref{eq:dfrac_reaction}) in the backward direction dominates the destruction (see Appendix \ref{appendix:opr_dh} for more details).
For example, the abundances of CO and electrons in  dense cloud conditions are typically 
$<$10$^{-4}$ and $\sim$10$^{-8}$-10$^{-7}$, respectively.
Then, even a small fraction of $\ohh$, $\sim$10$^{-3}$, can affect the deuterium chemistry at low temperatures.

In astrochemical simulations, it has been generally assumed that hydrogen (and deuterium) is already locked in 
H$_2$ (and HD) at the beginning of the simulations, and the focus is placed on the subsequent molecular evolution after H$_2$ (and HD) formation.
In this approach, the initial $\op{H_2}$ is treated as a free parameter, since the $\op{H_2}$ in the ISM is not well-constrained;
it requires detailed physical and chemical modeling to estimate the $\op{H_2}$ from observations of molecules other than H$_2$ 
in cold clouds/cores \citep[e.g.,][]{pagani09,pagani11,legal14}.
The resultant deuteration ratio of icy molecules in the simulations strongly depends on the initial $\op{H_2}$,
since the time it takes for the $\op{H_2}$ to reach steady state is longer \citep[several Myr,][]{flower06} than the timescale of
ice formation ($\sim$the freeze-out timescale) in dense cloud core conditions.

In this paper, we investigate the evolution of water deuteration and the $\op{H_2}$ 
during the chemical evolution from \mbox{\ion{H}{i}}-dominated clouds through the formation of denser molecular clouds.
One of the plausible scenarios of molecular cloud formation is that diffuse \mbox{\ion{H}{i}} gas is swept up 
by global accretion flows, such as supernova explosions, via expanding \mbox{\ion{H}{ii}} regions, or via colliding turbulent flows 
in the diffuse ISM \citep[e.g.,][]{hennebelle99,hartmann01,inoue12,inutsuka15}.
A recent review of molecular cloud formation can be found in \citet{dobbs14}.
To simulate this process, we revisit a one-dimensional shock model developed by \citet[][hereafter B04]{bergin04} and \citet[][hereafter H10]{hassel10} 
with post-processing gas-ice chemistry calculations.
This allows us to study $\op{H_2}$ and water deuteration during the formation and early evolution of molecular clouds 
without making an arbitrary assumption of the initial molecular abundances, including the initial $\op{H_2}$.
B04 studied H$_2$ and CO formation, while H10 focused on the formation of icy molecules without deuterium and nuclear spin-state chemistry. 

This paper is organized as follows: 
We describe our physical model and chemical model in Sections \ref{sec:phys} and \ref{sec:chem}, respectively.
In Section \ref{sec:result} we present the evolution of the $\op{H_2}$ and water ice deuteration in our fiducial model.
Parameter dependences are discussed in Section \ref{sec:parameter}, and 
comparisons between our and previous studies are made in Section \ref{sec:comparison}.
Our findings are summarized in Section \ref{sec:conclusion}.
In the Appendix, we derive some analytical formulae for the $\op{H_2}$, its effect on deuterium chemistry, and deuterium fractionation of water ice 
under irradiation conditions, which may help readers to understand the numerical results and the dependencies on adopted parameters.

\section{Physical model}
\label{sec:phys}
To simulate the formation and physical evolution of a molecular cloud, 
we use the one-dimensional shock model developed by B04 and H10.
Here we briefly outline the model, while more details can be found in the original papers.

The model describes the evolution of post-shock materials in a plane-parallel configuration.
In order to obtain density and gas temperature evolution, the conservation laws of mass and momentum are solved with the energy equation, 
considering time-dependent cooling/heating rates and simplified chemistry (B04).
The cosmic-ray ionization rate of H$_2$ ($\xi_{\rm H_2}$) is set to be $1.3 \times 10^{-17}$ s$^{-1}$.
Interstellar radiation is assumed to be incident from one side only, adopting the Draine field \citep{draine78}.
As time proceeds, the column density of post-shock materials increases, which assists molecular formation by absorbing interstellar radiation.
The column density at a given time $t$ after passing through the shock front is $N_{\rm H} = n_0v_0t$, 
where $n_0$ and $v_0$ are preshock \mbox{\ion{H}{i}} gas density and velocity of the accretion flow, respectively.
$N_{\rm H}$ is converted into $A_V$ using the relation $A_V/N_{\rm H} = 5\times 10^{-22}$ mag/cm$^{-2}$.
In this work we choose the model with $n_0 = 10$ cm$^{-3}$ and $v_0 = 15$ km s$^{-1}$ (model 4 in H10).
The preshock gas temperature is assumed to be 40 K.
The preshock density and temperature agree with observed properties of \mbox{\ion{H}{i}} gas in the ISM \citep{heiles03}.
The magnetic field in \mbox{\ion{H}{i}} clouds is of the order of $\mu$G \citep{heiles05}. 
In the scenario of molecular cloud formation via accretion of \mbox{\ion{H}{i}} gas, molecular clouds are formed in the 
regions where the angle between the shock normal and the mean magnetic field is sufficiently low; 
otherwise the magnetic pressure prevents the accumulation of the gas \citep{inoue09,inoue12}.
In our physical model, the magnetic field component parallel to the shock front is set to be 0.01 $\mu$G,
assuming that the shock normal and the mean magnetic field are almost parallel.
In this setting, the formation timescale of molecular clouds with $A_V=1$ mag is $\sim$4 Myr ($A_V$/1 mag)($n_0$/10 cm$^{-3}$)$^{-1}$ ($v_0$/15 km s$^{-1}$)$^{-1}$,
which is comparable to or less than the estimated lifetime of molecular clouds \citep{hartmann01, kawamura09}.
The impact of varying $n_0$ and $v_0$ is briefly discussed in Section \ref{sec:weaker_shock}.

\begin{figure}
\resizebox{\hsize}{!}{\includegraphics{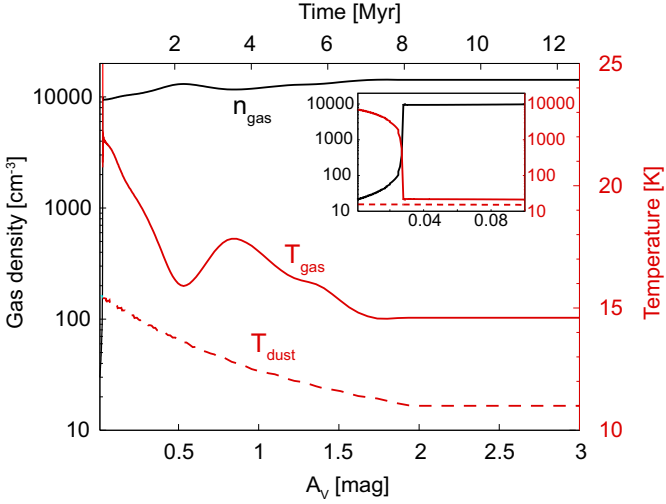}}
\caption{Physical evolution after passing through the shock front.
We adopt the same shock model as Model 4 of B04 and H10; 
i.e., the preshock density, temperature and velocity are 10 cm$^{-3}$, 40 K, and 15 km s$^{-1}$ respectively.
The inset shows the zoom-in view of the earliest evolution at $A_V \leq 0.1$ mag, 
which corresponds to $t \leq 0.3$ Myr.}
\label{fig:phys}
\end{figure}

Using density, gas temperature, and $A_V$ profiles, the dust temperature is calculated by equating the heating rate with the cooling rate.
Heating processes include heating by interstellar radiation and collisions with gas molecules, 
while cooling processes include thermal emission and sublimation of ices (H10).
Figure \ref{fig:phys} shows the evolution of gas density, gas temperature ($T_{\rm gas}$), and dust temperature ($T_{\rm dust}$) adopted in this work.

In the physical model, the effect of self-gravity is not considered.
Following the discussion by \citet{hartmann01}, internal pressure due to the self gravity of a sheet-like cloud exceeds ram pressure due to the accretion flow when
$A_V \gtrsim 4.5 \,\, {\rm mag} \,\, (P_{\rm e}/3.2\times10^5k_{\rm B})^{0.5}$, 
where $P_{\rm e}$ and $k_{\rm B}$ are the ram pressure of an accretion flow and the Boltzmann constant, respectively.
We use the shock model until $A_V$ reaches 3 mag, so that our simulation is in the regime where the ram pressure overwhelms the pressure due to self gravity.

\section{Chemical model}
\label{sec:chem}
To simulate the gas-ice chemistry during the formation and evolution of the cloud, we use a rate equation method adopting a three-phase model, 
which consists of gas, a chemically active icy surface, and an inactive inert ice mantle \citep{hasegawa93b}.
In three-phase models, it is often assumed that chemical processes occur only in the uppermost monolayer \citep[e.g.,][]{hasegawa93b,garrod11}.
However, the exact number of chemically active ice surface layers ($N_{\rm act}$) remains uncertain. 
Laboratory experiments have shown that atomic hydrogen can penetrate into the top several monolayers of ice, 
and reactions can occur in the ice as well as on the surface \citep{fuchs09,ioppolo10}.
\citet{vasyunin13a} find in their Monte-Carlo simulations that a model with $N_{\rm act} = 4$ can reproduce the 
experimental thermal desorption profiles of mixed ices better than a model with $N_{\rm act} = 1$.
We set $N_{\rm act}$ to be 4 in our fiducial model, 
so that the top four monolayers are chemically active and have uniform chemical composition.
In the rest of this paper, we refer to the outermost $N_{\rm act}$ layers 
as the active (surface) ice layers, and all of the layers including the active ice layers as the bulk ice mantle.

As chemical processes, we take into account gas-phase reactions, interactions between gas and (icy) grain surfaces, and surface reactions.
For photodissociation and photoionization in the gas phase, we consider the self-shielding of H$_2$, HD, CO, \mbox{\ion{C}{i}}, and N$_2$ 
and mutual shielding by H$_2$ of HD, CO, \mbox{\ion{C}{i}}, and N$_2$ \citep{draine98,kamp00,visser09,wolcott11,li13}.
Grain surface reactions are treated using the modified-rate equation method of \citet[][method A]{garrod08}.
It can take into account the competition between surface processes to improve the accuracy of the original rate equation method in the stochastic regime. 
%(i.e., when $\langle A \rangle \lesssim 1$ and $\langle B \rangle \lesssim 1$, where $\langle A \rangle$ is the population of species A on a surface).
Our chemical reaction network is originally based on the gas-ice reaction network of \citet{garrod06}.
When the gas temperature exceeds 100 K, we use the reaction sets in \citet{harada10,harada12} 
instead of those in \citet{garrod06} for the neutral-neutral reactions and neutral-ion reactions; 
the network of \citet{harada10,harada12} was designed for modeling high-temperature gas-phase chemistry (100-800 K) 
and includes reactions with a high potential energy barrier.
Our network has been extended to include mono, doubly, and triply deuterated species \citep{aikawa12b,furuya13} 
and nuclear spin states of H$_2$, H$_3^+$, and their isotopologues \citep[][]{hincelin14,coutens14b}.
The rate coefficients for the H$_2$ + H$_3^+$ system, including the isotopologues HD, D$_2$, H$_2$D$^+$, D$_2$H$^+$, and D$_3^+$ 
with their nuclear spin states, are taken from \citet{hugo09}.
In the rest of this sections, we describe the treatment of interaction between gas and (icy) grain surface, 
surface reactions (Section \ref{sec:surf}), and the initial abundance (Section \ref{sec:iniab}).

\subsection{Gas-grain interaction and surface chemistry}
\label{sec:surf}
When gaseous neutral species collide with a dust grain, they can stick to the surface.
The radius of a dust grain is set to be 0.1 $\mu$m with the dust-to-gas mass ratio of 0.01.
The sticking probabilities for H, D, H$_2$, HD, and D$_2$ are calculated as functions of gas and dust temperatures \citep{hollenbach79}, 
while those for heavier species are set to be unity.
Interactions between ions and dust grains are calculated in the same way as \citet{furuya12}; 
the collisional rates between ions and grains are calculated considering Coulomb focusing \citep{draine87}, and 
the products and branching ratios are assumed to be the same as the corresponding electron recombination in the gas phase.

Adsorbed species can desorb thermally into the gas phase again.
The rate of thermal desorption depends exponentially on the binding energy of absorbed species on a surface.
The binding energy of a species depends on the molecular composition of a surface.
For example, the binding energy of H$_2$ on an H$_2$O substrate is several hundred K \citep{hornekaer05}, while
that of H$_2$ on an H$_2$ substrate is much smaller \citep[23 K,][]{cuppen07}.  
In this work we use the binding energies on a water ice substrate, and
modify them depending on the surface coverage of H$_2$ \citep[$\theta_{\rm H_2}$,][]{garrod11}:
\begin{align}
E_{\rm des}(A) = (1-\theta_{\rm H_2})E_{\rm des}^0(A) + \theta_{\rm H_2} \left[E_{\rm des}^0(A) \times 
\left(\frac{23\,\,{\rm K}}{E_{\rm des}^0({\rm H_2})}\right)\right], \label{eq:edes}
\end{align}
where $E_{\rm des}^0(A)$ is the binding energy of species $A$ on a water substrate taken from \citet{garrod06}.
$E_{\rm des}^0({\rm H})$ and $E_{\rm des}^0({\rm H_2})$ are treated as free parameters, and we set them to be 550 K in our fiducial model.
Without the modification, a three phase model can give unphysical solutions;
in typical cold dark cloud conditions with $E_{\rm des}^0({\rm H_2})\gtrsim 400$ K, H$_2$ is depleted from the gas phase 
and thick pure H$_2$ ice mantles form \citep{garrod11}.
%Very recently, \citet{hincelin14a} proposed a new method to take into account the different binding energies 
%of H$_2$ on water and H$_2$ substrates.  
The desorption energy of atomic deuterium is set to be 21 K higher than that of atomic hydrogen 
($E_{\rm des}^0({\rm D}) = E_{\rm des}^0({\rm H}) + 21$ K), following \citet{caselli02}.
For other deuterated species, we use the same desorption energies as for normal species.

For non-thermal desorption processes, we consider stochastic heating by cosmic-rays \citep{hasegawa93a}, 
photodesorption \citep{westley95,oberg07}, and reactive desorption \citep{garrod07,dulieu13,vasyunin13b}.
We assume that roughly 1\% of products formed by exothermic reactions are desorbed following the method of \citet{garrod07},
except for the reactions OH + H $\rightarrow$ H$_2$O, and O + O $\rightarrow$ O$_2$.
Recently \citet{dulieu13} found that desorption probabilities of these reactions are very high, $>$90 \% and 60 \%, on a silicate substrate,
while the probabilities are much lower on an ice substrate.
\citet{minissale14} found that the desorption probability of O + O $\rightarrow$ O$_2$ becomes smaller with the increasing ice coverage.
Motivated by these experiments, we use the coverage dependent desorption probabilities for these two reactions and their deuterated analogues:
\begin{align}
P_{\rm reactdes} &= (1-\theta_{\rm ice})P_{\rm bare} + \theta_{\rm ice}P_{\rm ice}
\end{align}
where $\theta_{\rm ice}$ is the surface coverage of the ice.
$P_{\rm bare}$ is the desorption probability on a silicate substrate taken from \citet{dulieu13}, while $P_{\rm ice}$ is that on a ice substrate,
calculated by the method of \citet{garrod07}.

Photochemistry of ices is an essential process to prevent formation of thick ice mantles 
in gas with low extinction \citep[e.g.,][]{tielens05,cuppen07,hollenbach09}.
According to molecular dynamics (MD) simulations, there are several possible outcomes after a UV photon dissociates water ice;
(i) the photofragments are trapped on the surface, (ii) either of the fragments is desorbed into the gas phase, 
(iii) the fragments recombine and the product is either trapped on the surface or desorbed into the gas phase, etc \citep{andersson06,andersson08}.
The total photodissociation rates (cm$^{-3}$ s$^{-1}$) of water ice are calculated as follows \citep{furuya13,taquet13b}:
\begin{align}
R^{\rm tot}_{{\rm ph}, \,i} &= f_{{\rm abs}, \,i} \pi a^2 n_{\rm gr} F_{\rm UV} \exp(-\gamma_i A_V), \\
f_{{\rm abs}, \,i} &= \theta_i P_{{\rm abs}, \,i} \times {\rm min}(N_{\rm layer}, \,\, 4),
\end{align}
where $n_{\rm gr}$ is the number density of dust grains, 
$a$ is the radius of dust grains, 
$f_{{\rm abs}, \,i}$ is the fraction of the incident photons absorbed by species $i$ (i.e., water ice here), 
and $N_{\rm layer}$ is the number of monolayers of the bulk ice mantle.
$F_{\rm UV}$ is the flux of the FUV radiation field, 
assuming $2\times10^8$ photon cm$^{-2}$ s$^{-1}$ for interstellar radiation and $3\times$10$^3$ photon cm$^{-2}$ s$^{-1}$ for cosmic-ray induced UV radiation.
$P_{{\rm abs}, \,i}$ is the absorption probability of the incident FUV photon per monolayer.
We calculated $P_{{\rm abs}, \,i}$ by convolving wavelength dependent photoabsorption cross sections of 
water ice \citep{mason06} with the emission spectrum of the interstellar radiation
field \citep{draine78} and with that of the cosmic-ray-induced radiation field \citep{gredel89}.
When we calculate $f_{{\rm abs}, \,i}$, we consider the absorption by up to four outermost monolayers regardless of $N_{\rm act}$.
%{\bf Since we do not allow any chemistry occur in the inert ice mantle, 
%it is implicitly assumed that every photodissociation event in the inert ice mantle eventually leads to the recombination between the photofragments 
%(i.e. the recombination is assumed to be faster than any other reaction).}
The parameter $\gamma_i$ is for the attenuation of interstellar radiation field by dust grains, and we apply the value for photodissociation in the gas phase.
Recently, \citet{arasa15} performed MD simulations to study the possible outcomes of H$_2$O, HDO, and D$_2$O photodissociation in H$_2$O ice 
and their probabilities per dissociation ($b^j$) as functions of depth into the ice.
With their results, the rate of each outcome is calculated by $b^j R^{\rm tot}_{{\rm ph}, \,i}$.
%The outcomes and their probability per dissociation  are taken from XX and summarized in Table XX.
The total photodesorption yield of water ice (desorbed as OX or X$_2$O, X is H or D) per incident photon for thick pure water ice is 
$\sim$3$\times 10^{-4}$ in our model.

The absorption probability of the incident FUV photon for the other icy species are calculated 
as $P_{{\rm abs}, \,{\rm H_2O}} \times (k^0_{\rm ph, \,i}/k^0_{\rm ph, \,{\rm H_2O}})$,
where $k^0_{\rm ph}$ is the unattenuated photodissociation rate in the gas phase.
We also assume that all the photofragments are trapped in the active surface ice layers for simplicity.
Instead, the photodesorption rates are calculated as
\begin{align}
R_{{\rm phdes}, \,i} = \pi a^2 n_{\rm gr} F_{\rm UV} \exp(-\gamma_i A_V) \times \theta_i Y_i{\rm min}(N_{\rm layer}/4, \,\, 1),
\end{align}
where $Y_i$ is the photodesorption yield per incident FUV photon for thick ice ($N_{\rm layer} \geq 4$).
The term ${\rm min}(N_{\rm layer}/4, \,\, 1)$ is for thin ice ($N_{\rm layer} < 4$), assuming that the yield is linear to ice thickness.
We use the yields derived from experimental work for CO, CO$_2$, O$_2$, and N$_2$ pure ices \citep{oberg09,fayolle11,fayolle13}.
\citet{bertin12} found that photodesorption of CO is less efficient on a water substrate than on a CO substrate by more than a factor of 10.
Infrared observations suggest that water ice is the main constituent of interstellar ice in the early stage of ice formation, 
while CO is the dominant constituent of the outer layers of the ice \citep[e.g.,][]{oberg11}.
Considering the importance of CO abundance in the gas phase on deuterium chemistry, we treat the photodesorption yield of CO 
as a function of the surface coverage of CO:
\begin{align}
Y_{\rm CO} = (1-\theta_{\rm CO})Y_{\rm CO\mathchar`- H_2O} + \theta_{\rm CO}Y_{\rm CO\mathchar`- CO}, \label{eq:yco}
\end{align}
where $Y_{\rm CO\mathchar`- CO}$ is the photodesorption yield for pure CO ice taken from \citet[][10$^{-2}$]{fayolle11},
while $Y_{\rm CO\mathchar`- H_2O}$ is the photodesorption yield for CO interacting with H$_2$O.   
We assume $Y_{\rm CO\mathchar`- H_2O} = 3\times10^{-4}$.
In this way, the CO freeze-out is self-limited in our model; 
with increasing the surface coverage of CO, the photodesorption yield of CO increases and the CO freeze-out is delayed.
For species for which no data is available in the literature, we set $Y_i$ to be 10$^{-3}$.

\begin{figure}
\resizebox{\hsize}{!}{\includegraphics{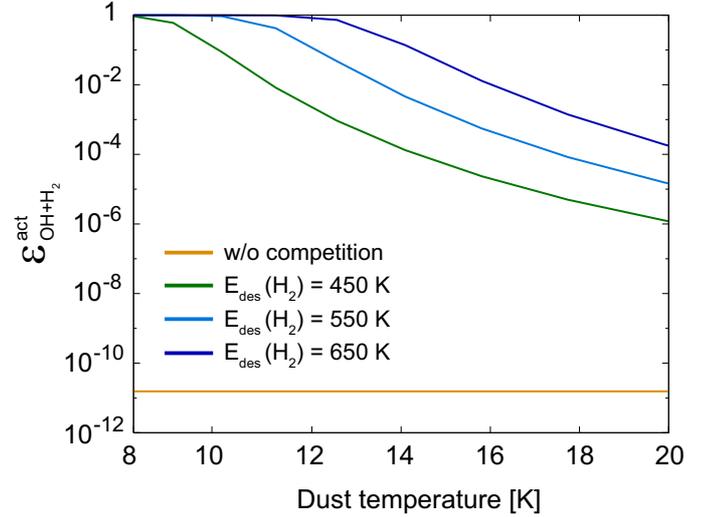}}
\caption{Efficiency factor for reaction OH + H$_2$ $\rightarrow$ H$_2$O + H, $\varepsilon^{\rm act}_{\rm OH+H_2}$, 
without (yellow) and with the reaction-diffusion competition in our models as functions of dust temperature.
For the case with the competition, $\varepsilon^{\rm act}_{\rm OH+H_2}$ with different $E_{\rm des}({\rm H_2})$ are shown by different colors.}
\label{fig:rdcomp}
\end{figure}

Surface reactions are assumed to occur by the Langmuir-Hinshelwood mechanism between physisorbed species;
adsorbed species diffuse by thermal hopping and react with each other when they meet \citep[e.g.,][]{hasegawa92}.
The reaction rate coefficient (s$^{-1}$) of surface reaction, A + B $\rightarrow$ AB, is given by
\begin{align}
%k_{A+B} &= \varepsilon^{\rm act}_{A+B} \times \frac{k_{\rm hop}(A) + k_{\rm hop}(B)}{N_{\rm act}N_{\rm S}}, \\
k_{A+B} &= \varepsilon^{\rm act}_{A+B} \times (k_{\rm hop}(A) + k_{\rm hop}(B))/(N_{\rm act}N_{\rm S}), \\
k_{\rm hop}(A) &= \nu_A \exp(-0.5 E_{\rm des}(A)/T_{\rm dust}), \label{eq:khop}
\end{align}
where $k_{\rm hop}(A)$ is the surface diffusion rate of species $A$ by thermal hopping from one binding site to another, 
and $N_{\rm S}$ is the number of binding sites on the surface per monolayer ($\sim2\times10^6$).
The energy barrier against hopping is set to be half of the desorption energy given in Equation (\ref{eq:edes}).
The pre-exponential factor $\nu_A$ is the vibrational frequency of species $A$ in the binding site
and is evaluated by using the harmonic oscillator strength \citep{hasegawa92}.
We do not allow the surface diffusion through quantum tunneling; 
laboratory experiments have shown that the thermal hopping is the dominant mechanism of 
the surface diffusion of atomic hydrogen on cold amorphous water ice \citep{watanabe10,hama12}.
The efficiency factor $\varepsilon^{\rm act}$ is for reactions with the activation energy barriers, 
otherwise it is set to unity.
When the reactants are in the same binding site, there would be the two possible outcomes;
a reaction occurs overcoming the activation energy barrier or either one of the reactants hop to another binding site 
before the reaction occurs \citep[][``reaction-diffusion competition"]{tielens82}.
This is naturally included in microscopic Monte Carlo simulations \citep[e.g.,][]{chang14}.
Considering the competition, $\varepsilon^{\rm act}$ is evaluated as follows \citep{tielens82,awad05,chang07}:
\begin{align}
\varepsilon^{\rm act}_{A+B} = \frac{\nu_{A+B}\kappa_{A+B}}{\nu_{A+B}\kappa_{A+B} + k_{\rm hop}(A) + k_{\rm hop}(B)}, \label{eq:rd_comp}
\end{align}
where $\nu_{A+B}$ is the collisional frequency of the reactants in the binding site and $\kappa_{A+B}$ is the probability 
to overcome the activation energy barrier per collision.
Assuming the collisional frequency can be approximated by the vibrational frequency in the binding site,
we set $\nu_{A+B}$ to be the largest of the vibrational frequency of either species A or B \citep{garrod11}.
The barriers are assumed to be overcome either thermally or via quantum tunneling, whichever is faster.
Transmission probabilities (probabilities of tunneling through the activation energy barriers) of reactions relevant for water ice are 
calculated assuming a rectangular potential barrier with a width of 1 \AA.
We treat the use of the reaction-diffusion competition as a free parameter in order to study its impact,
because most published rate equation models do not consider the reaction-diffusion competition.
When we do not consider the competition, Equation ($\ref{eq:rd_comp}$) is replaced by $\varepsilon^{\rm act}_{A+B} = \kappa_{A+B}$ 
as the analogy of gas phase reactions \citep[e.g.,][]{hasegawa92}.
Figure \ref{fig:rdcomp} shows the efficiency factor for reaction OH + H$_2$ $\rightarrow$ H$_2$O + H, $\varepsilon^{\rm act}_{\rm OH+H_2}$, 
with and without the competition in our models as functions of dust temperature.
The activation energy barrier of the reaction is set to be 2100 K \citep{atkinson04}, 
and thus it is overcome via quantum tunneling at low temperatures \citep[see][]{oba12}.
When the competition is considered, $\varepsilon^{\rm act}_{\rm OH+H_2}$ is extremely higher than the transmission probability.
Note that, when the competition is considered, $\varepsilon^{\rm act}_{\rm OH+H_2}$ decreases 
with increasing dust temperature and with decreasing $E_{\rm des}({\rm H_2})$, reflecting a increase of $k_{\rm hop}({\rm H_2})$.

We set the branching ratio for the H$_2$ formation on a surface ($\ohh$):($\phh$) to be 3:1 as experimentally demonstrated by \citet{watanabe10}.
They also find that the spin conversion occurs on amorphous water ice in a short timescale ($<$1 hour), 
while the mechanism remains uncertain.
We do not consider spin conversion on a surface in this work.

\subsection{Initial abundances and parameters}
\label{sec:iniab}
The amount of material in the ISM that is available for gas and ice chemistry remains largely uncertain \citep[e.g.,][]{sofia04}.
B04 and H10 used the elemental abundances based on the observations of the $\zeta$ Oph diffuse cloud.
We refer to elements heavier than oxygen as heavy metals in this paper.
\citet{greadel82} found that when they adopt much lower abundances of heavy metals than the $\zeta$ Oph values,
their gas-phase astrochemical model better reproduces the observation of various molecules in dense molecular clouds.
The elemental abundances similar to the $\zeta$ Oph diffuse cloud are referred to "high metal (HM)", 
while the reduced abundances are referred to "low metal (LM)" \citep{greadel82}.
Following Graedel et al., the LM abundances have been widely used in astrochemical models of dense molecular clouds.
The success of the LM abundances implies that the depletion of heavy metals from the gas phase occurs during the evolution 
from diffuse clouds to denser clouds, which is supported by some observational evidence \citep{joseph86}.
Plausible mechanisms of the depletion are thought to be entrapment in ices \citep{andersson13,kama15} 
and incorporation into refractory dust grains \citep{keller02}.
The former process is naturally included in our simulations, while the latter is not.

To study the impact of different initial abundances, 
we consider both the HM abundances and the LM abundances, which are summarized in Table \ref{table:element}.
All the elements are initially assumed to be in the form of either neutral atoms or atomic ions, depending on their ionization energy.
We use the LM abundances in our fiducial model.
%One may think that the use of the HM set is more appropriate for this work, 
%because at least the trapping of heavy metals in ice mantle is naturally included in our simulations.
%However, our knowledge of ice chemistry especially for heavy metals is rather limited at the present time; 
%because the dust temperature is $<15$ K during our simulations, partitioning of heavy metals between gas and ice is 
%controlled by non-thermal desorption processes, the rates of which are not well constrained.

%{\bf Polycyclic aromatic hydrocarbons (PAHs) are also important for determining ionization structure of gas, 
%because they can exchange charge with atomic ions on much shorter timescale than the electron recombination in the gas phase \citep[e.g.,][]{omont86,bakes98}.
%\citet{wakelam08} found that their gas-phase model with HM abundances well reproduces observational molecular abundances 
%in dense molecular clouds, if PAHs exits abundantly in the gas phase.
%PAHs in the gas phase have been detected in diffuse ISM \citep[][and references therein]{tielens08}, 
%but they have not been detected in dense molecular clouds, probably because PAHs are trapped in ice mantles as well as smaller molecules \citep{bouwman11}.
%We basically do not consider PAHs, while brief discussions on the impact of PAHs as charge carriers are given in Section \ref{sec:metal}.}

In Table \ref{table:parameters}, we summarize free chemical parameters we consider in this paper. 
The impact of the parameters is discussed in Section \ref{sec:parameter}.

\begin{table}
\caption{Initial abundances with respect to hydrogen nuclei.}
\label{table:element}
\centering
\begin{tabular}{ccc}
\hline\hline
Species & Low metal \footnotemark[1] & High metal \footnotemark[2] \\
\hline
H      & 1.00(0)    & 1.00(0)\\
D      & 1.50(-5)   & 1.50(-5)\\
He     & 9.75(-2)   & 9.75(-2)\\
C$^+$  & 7.86(-5)   & 7.86(-5)\\
N      & 2.47(-5)   & 2.47(-5)\\
O      & 1.80(-4)   & 1.80(-4)\\
Si$^+$ & 9.74(-9)   & 1.70(-6)\\
S$^+$  & 9.14(-8)   & 1.50(-5)\\
Fe$^+$ & 2.74(-9)   & 2.00(-7)\\
Na$^+$ & 2.25(-9)   & 2.00(-7)\\
Mg$^+$ & 1.09(-8)   & 2.40(-6)\\
Cl$^+$ & 2.16(-10)  & 1.17(-7)\\
P$^+$  & 1.00(-9)   & 1.80(-7)\\
\hline
\end{tabular}
\tablefoot{$a(-b)$ means $a\times10^{-b}$.\\
\tablefoottext{1}{The values are taken from \citet{aikawa99}. Deuterium abundance is from \citet{linsky03}.}
\tablefoottext{2}{Abundances for heavier elements than oxygen are taken from EA2 set in \citet{wakelam08}.}
}
%\tablenotetext{1}{$a(-b)$ means $a\times10^{-b}$.}
%\tablecomments{(a) $a(-b)$ means $a\times10^{-b}$.}
%% Any table notes must follow the \end{tabular} command.
\end{table}

\begin{table}
\caption{Summary of free chemical parameters.}
\label{table:parameters}
\centering
\begin{tabular}{cc}
\hline\hline
Parameters &  Values \footnotemark[1]\\
\hline
%$T^0_{\rm d}$ (K) & {\bf 15} - 20 \\
$E^0_{\rm des}$(H), $E^0_{\rm des}$(H$_2$) (K) & 450 - {\bf 550} - 650 \\
Reaction-diffusion competition? & {\bf Yes} - No \\
Elemental abundances & {\bf Low metal} - High metal\\
$N_{\rm act}$ & 1 - {\bf 4} \\
\hline
\end{tabular}
\tablefoottext{1}{Values used in our fiducial model are shown in bold letters.}
\end{table}
 
\section{Results from the fiducial model}
\label{sec:result}
\subsection{Overview of molecular formation}
\label{sec:overview}

\begin{figure}
\resizebox{\hsize}{!}{\includegraphics{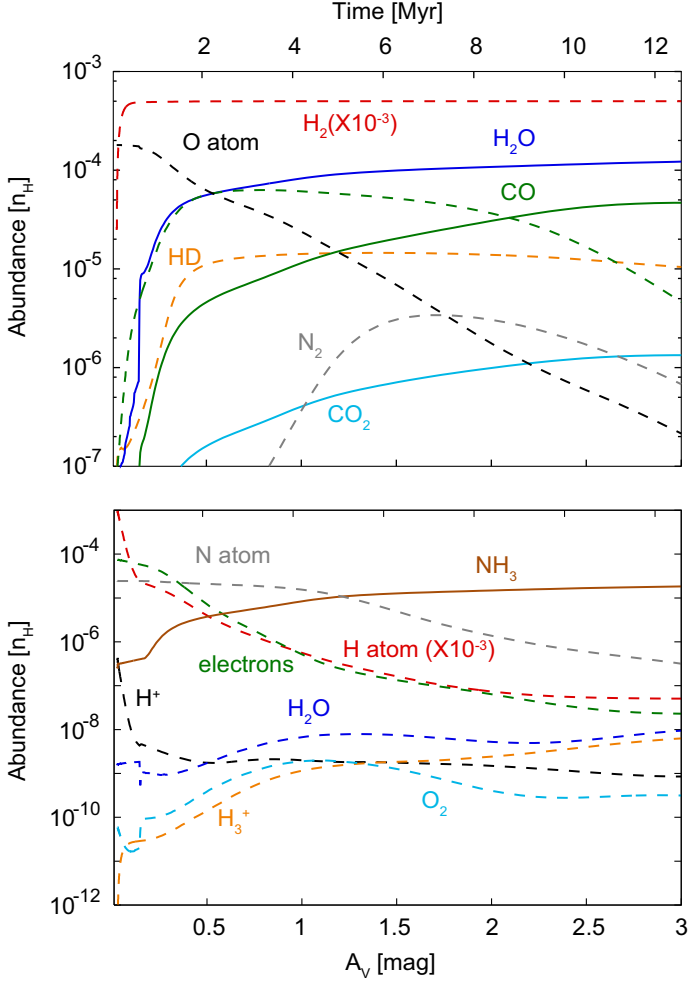}}
\caption{Fractional abundances of selected species with respect to hydrogen nuclei in the fiducial model as functions of visual extinction, or time.
The solid lines represent species in the bulk ice mantle, while the dashed lines represent gaseous species.
The H$_2$ and H atom abundances are multiplied by a factor of 10$^{-3}$.}
\label{fig:ab}
\end{figure}

Figure \ref{fig:ab} shows the fractional abundances of selected species as functions of visual extinction.
The horizontal axis is also read as time after passing through the shock front (see the labels at the top).
Note that $A_V$ in our model is likely different from the line of site visual extinction estimated from observations;
in reality, molecular clouds can be highly structured and UV photons can penetrate into the clouds along a direction where extinction is lower 
than the line of site visual extinction, while in our model interstellar radiation is assumed to be 
incident from one side only following the plane of the shock.
%{\bf According to three-dimensional hydrodynamics simulations of the gravitational evolution of turbulent (molecular) clouds \citep{clark14}, 
%even if the line of site visual extinction toward their clouds is greater than 10 mag, 
%the majority of gas on the line of site has the angle-averaged visual extinction of a few mag or less.}

In the shock front, the gas is heated up to 7000 K. 
As the gas is cooled via the line emission of \mbox{\ion{O}{i}} and \mbox{\ion{C}{ii}} and the gas density increases 
from 10 cm$^{-3}$ to 10$^4$ cm$^{-3}$, the conversion of atomic hydrogen to molecular hydrogen occurs via grain surface reactions, 
supported by the self-shielding of UV radiation.
Following H$_2$ formation, other self-shielding species with smaller abundances, CO and HD, become abundant (B04).
Another self-shielding species N$_2$ becomes moderately abundant later ($A_V \sim 1.5$ mag), though
N$_2$ is not the primary reservoir of nitrogen during the simulation, in agreement with previous studies \citep{maret06,daranlot12}. 
At $A_V \lesssim 1$ mag the primary reservoir is atomic nitrogen, while it is ammonia ice in the later times.

When the formation rate of water ice exceeds the photodesorption rate, dust grains start to be covered with water ice.
Thick ice mantles are already formed at $A_V = 0.5$ mag, and the water ice abundance reaches 10$^{-4}$ at $A_V \sim 1.5$ mag.
The abundance of atomic oxygen in the gas phase is already less than 10$^{-5}$ at $A_V \sim 1.5$ mag, which corresponds to $\sim$5 \% of elemental oxygen.
%{\bf There are some observations which show that the \mbox{\ion{O}{i}}/CO column density ratio 
%in molecular clouds is much larger than unity \citep{caux99,vastel00}.}
The main surface formation pathways of water ice in our model are as follows:
\begin{align}
{\rm O} + {\rm H} &\rightarrow {\rm OH},\\
{\rm OH} + {\rm H} &\rightarrow {\rm H_2O}, \label{react:oh+h} \\
{\rm OH} + {\rm H_2} &\rightarrow {\rm H_2O} + {\rm H}. \label{react:oh+h2}
\end{align}
While the OH + H$_2$ reaction in the gas phase has an activation energy barrier of 2100 K \citep{atkinson04},
\citet{oba12} experimentally demonstrated that Reaction (\ref{react:oh+h2}) occurs on a cold substrate of 10 K via quantum tunneling.
We used the barrier height of 2100 K to calculate the transmission probability of Reaction (\ref{react:oh+h2}), 
assuming a rectangular potential barrier with a width of 1 \AA.
\citet{taquet13b} calculated the transmission probability taking into account the shape of the barrier with the Eckart model.
The probability calculated by \citet{taquet13b} is $4\times10^{-7}$, which is larger than that in our model by four orders of magnitude.
It is likely therefore that we underestimate the rate coefficient of Reaction (\ref{react:oh+h2}).

In our fiducial model, Reaction (\ref{react:oh+h}) is the most efficient to form water ice 
at $A_V \lesssim 0.2$ mag at which $T_{\rm dust} \gtrsim 14$ K and $\num{H}/\num{H_2} \gtrsim 10^{-2}$, where $n(i)$ is the number density of species $i$.
As $A_V$ increases, the dust temperature and H/H$_2$ abundance ratio decrease.
A reduced dust temperature leads to a longer duration time of H$_2$ in a binding site, 
and thus to a higher reaction efficiency (see Figure \ref{fig:rdcomp});
the relative contribution of Reaction (\ref{react:oh+h2}) to water ice formation increases with increasing $A_V$.
Reaction (\ref{react:oh+h2}) then dominates over Reaction (\ref{react:oh+h}) at $A_V \gtrsim 0.2$ mag.
\citet{cuppen07} found that Reaction (\ref{react:oh+h}) is the main water ice formation pathway in diffuse and translucent cloud conditions using
microscopic Monte-Carlo simulations, while they found Reaction (\ref{react:oh+h2}) to be the more important in colder and denser conditions.
Our result is qualitatively consistent with their findings.
We have confirmed that in our model without the reaction-diffusion competition, Reaction (\ref{react:oh+h}) always dominates over Reaction (\ref{react:oh+h2}) 
during the simulation.
As discussed in Section \ref{sec:dh}, water ice deuteration heavily depends on which formation path, Reaction (\ref{react:oh+h}) or (\ref{react:oh+h2}),
is more effective.

Water ice can also be formed via the sequential hydrogenation of O$_2$ \citep{miyauchi08,ioppolo08}.
This pathway is not important in our model, because the accretion rate of atomic oxygen onto the icy grain surface is 
lower than that of atomic hydrogen by two orders of magnitude during the simulation.
Adsorbed atomic oxygen likely reacts with atomic hydrogen before another atomic oxygen accretes onto the surface. 

The O$_2$ abundance in the gas phase is as low as 10$^{-9}$ during the simulation. 
Most of the oxygen is locked in either water ice or CO before $A_V$ reaches 1.5 mag.
O$_2$ is efficiently destroyed by photodissociation at $A_V < 1.5$ mag, 
while the amount of atomic oxygen in the gas phase that is available for O$_2$ formation is limited at $A_V > 1.5$ mag.
The model result is essentially the same as the scenario proposed by \citet{bergin00} 
to explain the low upper limits on the O$_2$ abundance ($<10^{-7}$) observationally derived in nearby clouds \citep{goldsmith00,pagani03}.

\begin{figure}
\resizebox{\hsize}{!}{\includegraphics{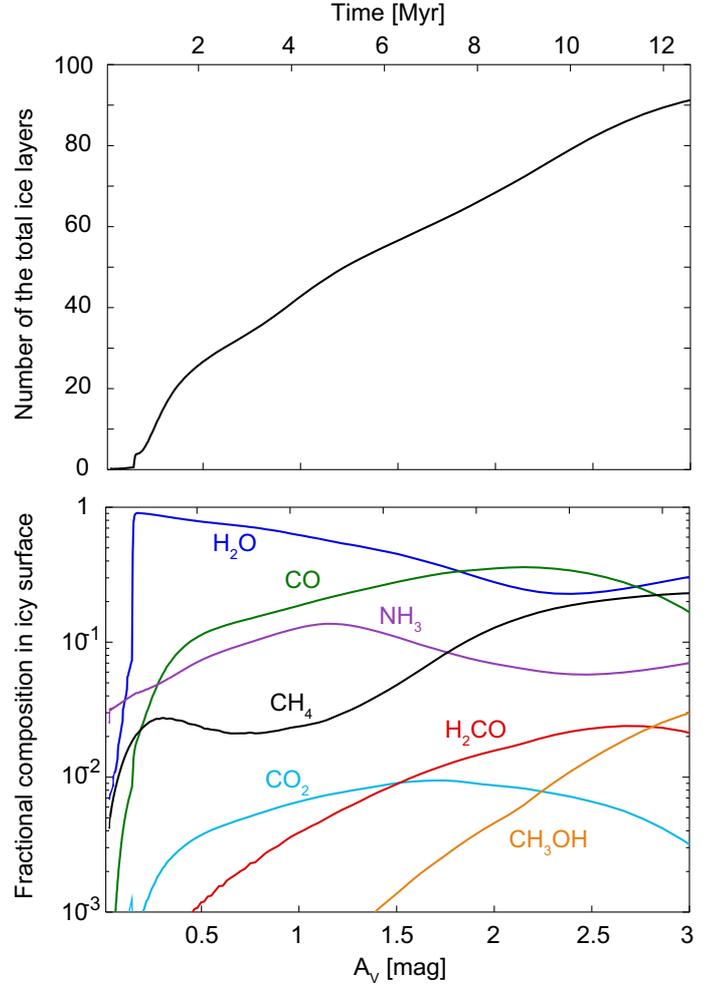}}
\caption{The number of icy layers in total (top) and fractional compositions in the active surface ice layers (bottom) 
as functions of visual extinction, or time.}
\label{fig:layer}
\end{figure}

Figure \ref{fig:layer} shows the number of icy layers in total and the fractional composition in the icy surface layers as functions of $A_V$.
The total number of ice layers formed during the simulation is $\sim$90 monolayers.
At $A_V \lesssim 1.5$ mag ($N_{\rm layer} \lesssim 60$), the surface layers of ice mantle is predominantly composed of water, 
while in the later stage, other molecules, such as CO, become as abundant as water.
%We ran the fiducial model with the constant CO photodesorption yield of 10$^{-2}$ and $3\times10^{-4}$ 
%to check the impact of the use of the coverage dependent yield.}

The ice composition at $A_V$ = 3 mag in our model is H$_2$O:CO:CO$_2$:CH$_4$:NH$_3$=100:38:1:15:15, 
which is compared with the observed median ice composition toward low-mass (high-mass) 
protostellar sources, 100:29(13):29(13):5(2):5(5) \citep{oberg11}.
The largest discrepancy is in CO$_2$.
The CO$_2$/H$_2$O ratio in our model is about one order of magnitude lower than the observed ratio.
We may overestimate the barrier against hopping for CO on the ice surface \citep{garrod11}. 
Since CO$_2$ ice formation (e.g., CO + OH $\rightarrow$ CO$_2$ + H) competes with water ice formation, our model may overestimate water ice formation.
Our model overestimates the CH$_4$/H$_2$O and NH$_3$/H$_2$O ratios by a factor of three or more.
Current gas-ice chemical models overestimate these two ratios in general \citep{garrod11,vasyunin13a,chang14}.
The reason remains unclear.

%{\bf We confirmed that the use of the reduced energy barrier against hopping (0.4 times the desorption energy) increases the CO$_2$/H$_2$O ice ratio 
%by a factor of around 2, although the model still underestimates the CO$_2$/H$_2$O ratio.} 

\subsection{Ortho-to-para nuclear spin ratio of H$_2$}
\label{sec:oprh2}
$\op{H_2}$ is determined by the competition between H$_2$ formation on grain surfaces and ortho-para spin conversion in the gas phase 
through proton exchange reactions with H$^+$ and H$_3^+$ \citep{gerlich90,lebourlot91,honvault11}.

If the timescales of these reactions are shorter than that of temporal variation of physical conditions, 
the $\op{H_2}$ reaches the steady state value,
\begin{align}
\opst{H_2} &= \frac{\beta_1 + b_{\rm o}\beta_2}{1 + (1-b_{\rm o})\beta_2}, \label{eq:oprsteady}\\
\beta_1 &=\tau_{\rm o\rightarrow p}/\tau_{\rm p\rightarrow o}, \label{eq:beta1}\\
\beta_2 &= \tau_{\rm o\rightarrow p}/\tau_{\rm H_2}, \label{eq:beta2}
\end{align}
where $b_o$ is the branching ratio to form $\ohh$ for H$_2$ formation on grain surfaces \citep[0.75,][]{watanabe10}.
$\tau_{\rm o \rightarrow p}$ and $\tau_{\rm p \rightarrow o}$ are the spin conversion timescale between $\ohh$ and $\phh$ through
the proton exchange reactions, while $\tau_{\rm H_2}$ is the formation timescale of H$_2$ on grain surfaces.
In the case with $\tau_{\rm H_2} \ll \tau_{\rm o \rightarrow p} \lesssim \tau_{\rm p \rightarrow o}$ (or $\beta_2 \gg 1$), $\opst{H_2}$ is $\sim b_{o}/(1-b_o) = 3$, 
i.e., the statistical value.
In another extreme case, $\tau_{\rm H_2} \gg \tau_{\rm o \rightarrow p}$, $\opst{H_2}$ is $\sim (\beta_1 + b_{o}\beta_2)$, 
i.e., larger than the thermalized value $\beta_1$.
The derivation of Equation (\ref{eq:oprsteady}) can be found in Appendix \ref{appendix:h2_opr} \citep[see also][]{lebourlot91}.

\begin{figure}
\resizebox{\hsize}{!}{\includegraphics{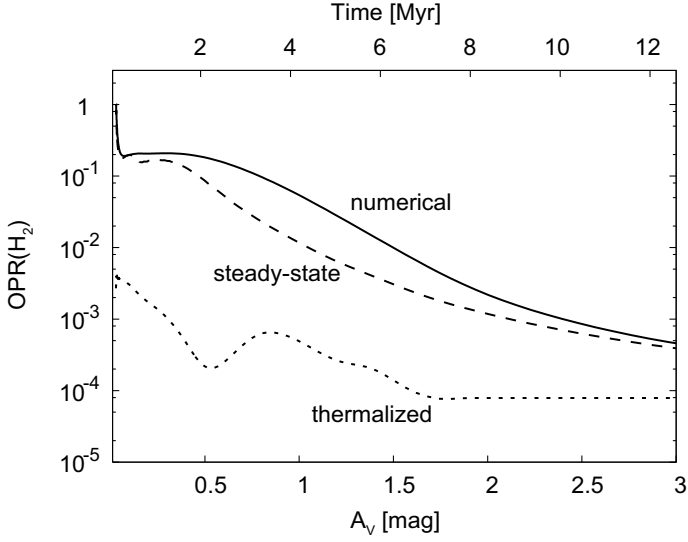}}
\caption{$\op{H_2}$ in the fiducial model as a function of visual extinction. The dashed line shows the steady-state value 
given by Equation (\ref{eq:oprsteady}) using molecular abundances in the simulations, 
while the dotted line shows the low-temperature thermalized value, $9\exp(-170.5/T_{\rm gas})$.}
\label{fig:opr}
\end{figure}

Figure \ref{fig:opr} shows the $\op{H_2}$ as a function of $A_V$ in the numerical simulation.
The thermalized value ($\beta_1 = 9\exp(-170.5/T_{\rm gas})$) and the steady-state value given by Equation (\ref{eq:oprsteady}) are also shown.
Since our physical model is time-dependent, the steady-state value can be calculated at given time.
The expression of the thermalized value is valid at $T_{\rm gas} \lesssim 100$ K, 
where the population of higher rotational levels ($J \geq 2$) is negligible.
When most hydrogen is locked in molecular hydrogen, the $\op{H_2}$ is already much less than the statistical value of three 
via ortho-para spin conversion in the gas phase, while it is much higher than the thermalized value.
The $\op{H_2}$ decreases with increasing $A_V$, approaching the steady state value, and reaches $\sim 5\times10^{-4}$ at $A_V = 3$ mag.
Note that the steady state value itself depends on $A_V$, reflecting local physical and chemical conditions.   
While the $\op{H_2}$ evolves in a non-equilibrium manner, the difference between the numerical and steady-state values is 
at most a factor of five during the simulation.
More details of the $\op{H_2}$ evolution are discussed below.

In the H/H$_2$ transition region at $\lesssim$0.1 mag, ortho-para spin conversion occurs predominantly through the following reactions:
\begin{align}
\reacteq{\ohh}{H^+}{\phh}{H^+} + 170.5\,\,{\rm K}. \label{react:h2+h+}
\end{align}
Assuming that the gas temperature is 20 K and the H$_2$ formation rate is half of the accretion rate of atomic hydrogen onto dust grains, 
we can evaluate $\opst{H_2}$ in the H/H$_2$ transition region by following (see Appendix \ref{appendix:h2_opr_transition} for the derivation):
\begin{align}
\begin{split}
\opst{H_2} \approx \beta_2 \approx &0.2 \bb{\frac{\ab{H_2}}{0.5}}^{-1}\bb{\frac{\xi_{\rm H_2}/n_{\rm H}}{8\times 10^{-22} \,\, {\rm cm^3 \,\, s^{-1}}}}^{-1} \\ 
                                    &\times \bb{\frac{\pi a^2 x_{\rm gr}}{4\times 10^{-22} \,\, {\rm cm^{2}}}}, \label{eq:opr_atomic}
\end{split}
\end{align}
where $\ab{H_2}$ and $x_{\rm gr}$ are the abundances of H$_2$ and dust grains, respectively, with respect to hydrogen nuclei, and
$\xi_{\rm H_2}/n_{\rm H}$ is the cosmic-ray ionization rate of H$_2$ divided by the number density of hydrogen nuclei.
Since $\beta_2 < 1$ is equivalent to $\tau_{\rm o \rightarrow p} < \tau_{\rm H_2}$, 
$\op{H_2}$ is almost in the steady state in the H/H$_2$ transition region as seen in Figure \ref{fig:opr}.
The equation reproduces the trend in the simulations that $\op{H_2}$ decreases with increasing H$_2$ abundance, 
and reproduces the numerical value of $\sim$0.2 when the conversion almost finishes.

In this work, we adopt a cosmic-ray ionization rate of $1.3\times10^{-17}$ s$^{-1}$, 
which is appropriate for dense cloud cores \citep{caselli98} and was used in B04.
However, that is an order of magnitude lower than the values now thought to be appropriate for interstellar gas 
with low extinction \citep{lepetit04,indriolo12}.
If we adopt the higher value, which would be more appropriate for the H/H$_2$ transition region, 
the $\op{H_2}$ would be as low as 10$^{-2}$ when the conversion of hydrogen into H$_2$ almost finishes.

After the H/H$_2$ transition, the abundance of atomic hydrogen continues to decrease toward the value 
$\num{H} \sim 1 \,\, {\rm cm^{-3}} \,\,(\xi_{\rm H_2}/10^{-17} \,\, {\rm s^{-1}})$ \citep[e.g.,][]{tielens05}.
The H$^+$ abundance stays nearly constant at $A_V \gtrsim 0.5$ mag, while H$_3^+$ abundance increases with increasing $A_V$, 
because of CO freeze-out and the drop of the ionization degree;
so that the reaction of H$_3^+$ with $\ohh$ dominates over Reaction (\ref{react:h2+h+}) in the forward direction at $A_V \gtrsim 1.5$ mag.
In summary, as $A_V$ increases, the abundance of atomic hydrogen decreases leading to an increase of $\tau_{\rm H_2}$, 
while the abundances of the light ions stay constant or increase  leading to an decrease of $\tau_{\rm o \rightarrow p}$.
The different behaviors of a light ions and atomic hydrogen lead to a decrease of the $\op{H_2}$ with increasing $A_V$.

\subsection{Water ice deuteration}
\label{sec:dh}
\begin{figure}
\resizebox{\hsize}{!}{\includegraphics{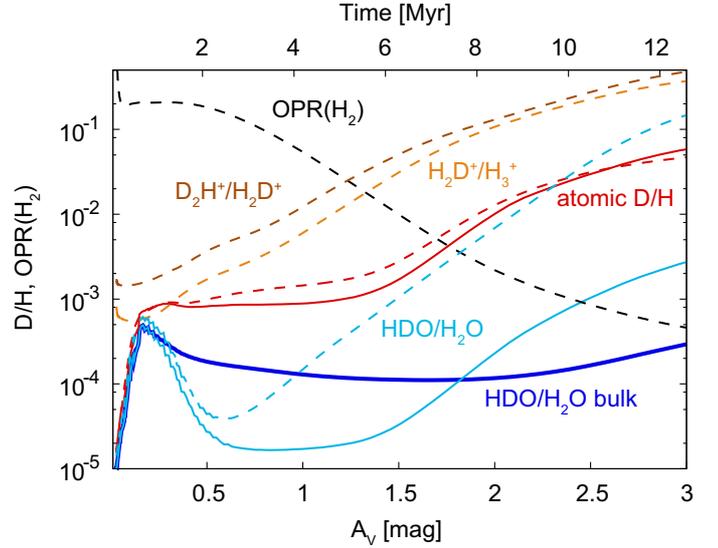}}
\caption{Deuteration of selected species and $\op{H_2}$ in the fiducial model as functions of visual extinction, or time.
The solid lines represent icy species in the active surface ice layers, while the dashed lines represent gaseous species.
The $\hdo$ ratio in the bulk ice mantle is shown by the thick solid line.}
\label{fig:dh}
\end{figure}

Figure \ref{fig:dh} shows the deuteration ratio of selected species as functions of $A_V$.
The H$_2$D$^+$/H$_3^+$ and D$_2^+$H/H$_2$D$^+$ ratios increase with increasing $A_V$, simply reflecting the decrease of the $\op{H_2}$ 
and the freeze-out of CO.
Although H$_3^+$ is highly enriched in deuterium, non-deuterated H$_3^+$ is still more abundant than its isotopologues.
The atomic D/H and icy HDO/H$_2$O ratios show more complicated behaviors.
The $\hdo$ ratio in the bulk ice mantle preserves the past physical and chemical conditions which the materials experienced, 
while the $\hdo$ ratio in the active ice surface layers reflects local physical and chemical conditions.
Interestingly, the latter ratio is less than the atomic D/H ratio by an order of magnitude 
except for the very early stage, where the two ratios are similar (see below for the reason).
At $A_V = 3$ mag, the $\hdo$ ratio in the bulk ice mantle is $\sim$2$\times$10$^{-4}$, 
while that in the active surface ice layers is $\sim$3$\times$10$^{-3}$.
The $\hdo$ ratio in the bulk ice mantle is so low that methane and ammonia are the major deuterium reservoirs in the ice; 
their D/H ratios are $\sim$10$^{-2}$, while their abundances with respect to water are $\gtrsim$0.1.
Compared to the median composition of interstellar ices, our model overestimates the abundances of methane and ammonia with 
respect to water by a factor of three or more as noted in Section \ref{sec:overview}.
Even considering this discrepancy, water is not the dominant deuterium reservoir in ice in our fiducial model.
We note that most deuterium is in the gas phase as HD or atomic deuterium during the simulation,
and only 4 \% of deuterium is in ice at most.

\subsubsection{The role of the ortho-to-para ratio of H$_2$}
\begin{figure}
\resizebox{\hsize}{!}{\includegraphics{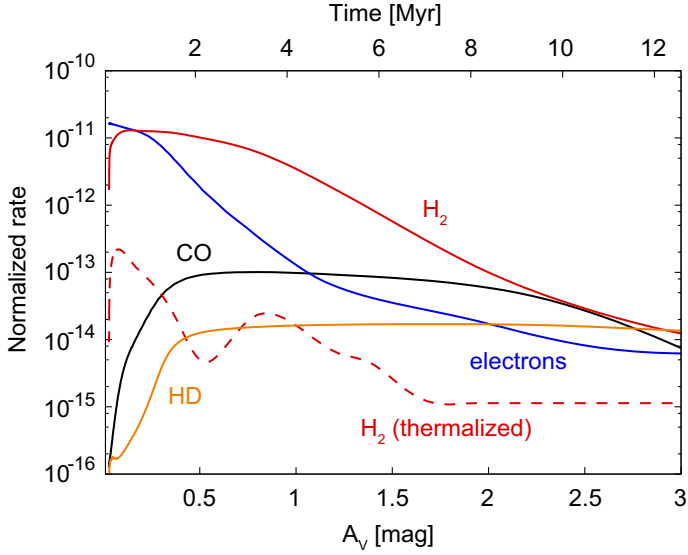}}
\caption{Normalized destruction rate of H$_2$D$^+$ by the reactions with CO and HD, recombination with electrons, and
the backward reaction with H$_2$ in the fiducial model as functions of $A_V$. 
The normalized destruction rates by H$_2$ assuming the thermalized ortho to para ratio of H$_2$ are also shown for comparison.}
\label{fig:dest_rate}
\end{figure}

Figure \ref{fig:dest_rate} shows the normalized destruction rates of H$_2$D$^+$ via the reactions 
H$_2$D$^+$ + CO, H$_2$D$^+$ + e, H$_2$D$^+$ + H$_2$, and H$_2$D$^+$ + HD as functions of visual extinction.
It shows that the main destroyer of H$_2$D$^+$ is H$_2$ except at the lowest visual extinction during the simulation.
Nevertheless, the effect of the $\op{H_2}$ on ice deuteration is minor in the fiducial model; 
we confirmed that the difference in the $\hdo$ ratio in the bulk ice at $A_V = 3$ mag between the models 
with and without nuclear spin chemistry is less than a factor of two.
This is mainly because the dominant production pathway of atomic deuterium is photodissociation of HD at $A_V \lesssim 1.5$ mag, 
during which water ice is mainly formed.
The contribution of D$_2$ photodissociation is minor because of its low abundance.
Only after the majority of elemental oxygen is locked up in water and CO does the dominant production pathway of atomic deuterium 
become recombination of deuterated ions with electrons.
In addition, at $A_V \gtrsim 1.5$ mag, the destruction rate of H$_2$D$^+$ by H$_2$ is comparable to that by CO. 

\subsubsection{The role of water ice photodissociation}
In gas with low extinction, water ice is efficiently photodissociated by interstellar UV photons followed by reformation via surface reactions.
The photodesorption yield of water (desorbed as OX or X$_2$O, where X is H or D) per photodissociation is only a few \% 
in the uppermost monolayer of the ice mantle, and the yield is lower in the deeper layers \citep{andersson06}.
When photodesorption regulates the growth of ice mantle, icy molecules on the surface experience the dissociation-reformation cycle 
many times before being buried in the inert ice mantle.
We find that this photodissociation-reformation cycle can cause the much lower $\hdo$ ratio in the active surface ice layers ($f_{\rm HDO,\,s}$) 
than the atomic D/H ratio ($f_{\rm D,\,s}$) as discussed below.
In our fiducial case, $f_{\rm HDO,\,s}$ is $\lesssim$10$^{-3}$ during the simulation, 
which ensures that the resultant $\hdo$ ratio in the bulk ice is $<$10$^{-3}$, 
although $f_{\rm D,\,s}$ reaches $>10^{-2}$ at the maximum.

\begin{figure}
\resizebox{\hsize}{!}{\includegraphics{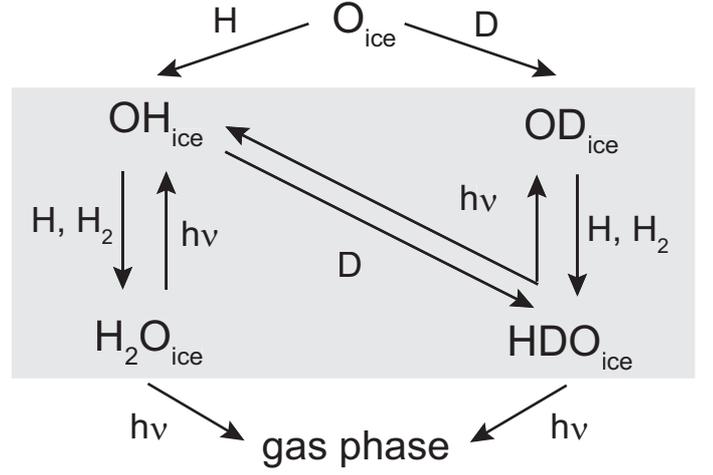}}
\caption{Important reaction routes to form H$_2$O and HDO on the (icy) grain surface in our models.
The gray area indicates the cycle of photodissociation and reformation of water ice, which is
discussed in the text.}
\label{fig:dh_network}
\end{figure}

Figure \ref{fig:dh_network} summarizes the important reactions for H$_2$O ice and HDO ice in our models.
Photodissociation of HDO ice has two branches, (${\rm OD} + {\rm H}$) and (${\rm OH} + {\rm D}$) with a branching ratio of 2:1 \citep{koning13}.
If most OH ice photofragments are converted to H$_2$O ice, 
HDO ice photodissociation leads to the removal of deuterium from the water ice chemical network on the timescale of photodissociation.
The efficiency of the deuterium removal depends on the probability for the OH ice photofragment to be converted back to HDO ice ($\phdo$).
$\phdo$ depends on the main formation pathway of H$_2$O, OH + H$_2$ $\rightarrow$ H$_2$O + H or OH + H $\rightarrow$ H$_2$O.
The conversion of OH into HDO would mainly occur via OH + D regardless of the main formation pathway of H$_2$O, which is confirmed in our models.
There are two reasons:
(i) OH + HD $\rightarrow$ HDO + H is much less efficient than OH + HD $\rightarrow$ H$_2$O + D owing to the mass dependency of tunneling reactions \citep{oba12}, 
and (ii) the HD/H$_2$ ratio is usually much less than the atomic D/H ratio.
Then, $\phdo$ becomes smaller, i.e., the deuterium removal becomes more efficient, with an increase in the contribution of the OH+H$_2$ pathway to the H$_2$O formation.

We can analytically evaluate the steady-state value of $f_{\rm HDO,\,s}$ with respect to $f_{\rm D,\,s}$ established 
through the photodissociation-reformation cycle (see Appendix \ref{appendix:water_dh} for details).
Figure \ref{fig:waterdh_analytic} shows the steady-state value of $f_{\rm HDO,\,s}$ as a function of $\Gamma \equiv R_{\rm OH+H_2}/R_{\rm OH+H}$, 
where $R_{\rm A+B}$ is the rate of the surface reaction A + B.
It demonstrates that when $\Gamma \gg 1$, $f_{\rm HDO,\,s}$ can be smaller than $f_{\rm D,\,s}$ by more than one order of magnitude.
As mentioned in Section \ref{sec:overview}, in our fiducial model, 
the OH+H pathway dominates over the OH+H$_2$ pathway at $A_V \lesssim 0.2$ mag, 
while the OH+H$_2$ pathway is dominant at later times.
Reflecting the change of the main formation pathway, $f_{\rm HDO,\,s}$ is similar to $f_{\rm D,\,s}$ at $A_V \lesssim 0.2$ mag, 
while $f_{\rm HDO,\,s}$ is much smaller than $f_{\rm D,\,s}$ at later times.

The analogous mechanism is not at work for methane and ammonia ices in our model, 
because the reactions $\react{CH_3}{H_2}{CH_4}{H}$ and $\react{NH_2}{H_2}{NH_3}{H}$ are much less efficient than the hydrogenation of CH$_3$ and NH$_2$ 
via reactions with atomic hydrogen;
the reactions with H$_2$ have high activation energy barriers of $\sim$6000 K \citep{kurylo69,aannestad73,hasegawa93a}.
Then, methane and ammonia ices are much more enriched in deuterium compared to water ice in our fiducial model.

\begin{figure}
\resizebox{\hsize}{!}{\includegraphics{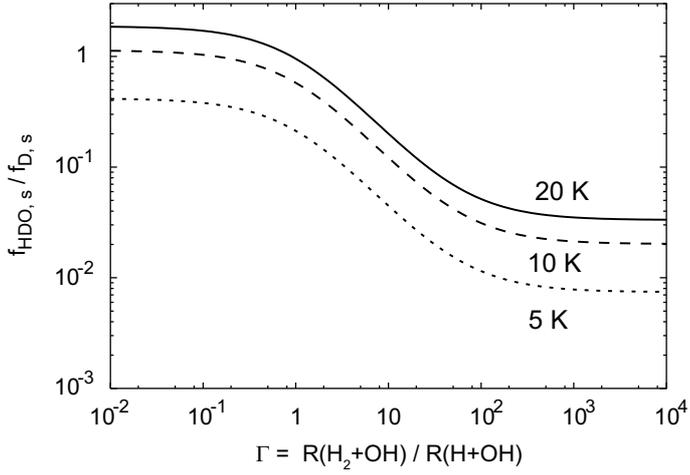}}
\caption{Steady-state $f_{\rm HDO, \, s}/f_{\rm D, \, s}$ ratio in the active ice layers as a function of the ratio of the reaction rate between 
OH + H$_2$ $\rightarrow$ H$_2$O + H and OH + H $\rightarrow$ H$_2$O at three temperatures given by Equation (\ref{eq:fhdo_fd_final}).
See Appendix \ref{appendix:water_dh} for more details.}
\label{fig:waterdh_analytic}
\end{figure}

\subsection{Water vapor deuteration}
\label{sec:dh_gas}
The abundance of water vapor lies in the range of (1-10)$\times10^{-9}$ during most of the simulation, as shown in Figure \ref{fig:ab}.
The $\hdo$ ratio in the gas phase is shown in Figure \ref{fig:dh}.
At $A_V \lesssim 0.5$ mag, $\hdo$ in the gas phase and that in the active surface ice layers are very similar; 
both H$_2$O and HDO vapors are formed by photodesorption and destroyed by reaction with C$^+$ to form HCO$^+$ or DCO$^+$,
followed by dissociative electronic recombination to form CO.
At $A_V \gtrsim 0.5$ mag, $\hdo$ ratio in the gas phase is much higher than that in the active surface ice layers;
the H$_2$O vapor abundance is mostly determined by the balance 
between photodoesorption and photodissociation in the gas phase at $A_V \lesssim 2.5$ mag \citep{hollenbach09},
while HDO is mainly formed via the sequential ion-neutral reactions, starting from $\react{O}{H_2D^+}{OD^+}{H_2}$, and destroyed by photodissociation.
For example, at $A_V = 1.5$ mag, the gas phase formation rate of HDO is faster than the photodesorption rate of HDO ice by a factor of 30, 
which corresponds to the difference between the $\hdo$ ratio in the gas phase and that in the active ice layers.   
At $A_V \gtrsim 2.5$ mag, both H$_2$O and HDO are formed via gas phase reactions and destroyed via photodissociation.
Then, there is no relation between the ratios in the gas and in the active ice layers.
The contribution of the photodesorption decreases as the UV field is attenuated and 
the fraction of water in the active ice layers decreases as shown in Figure \ref{fig:layer}.

\section{Parameter dependence}
\label{sec:parameter}
\subsection{Elemental abundances of heavy metals}
\label{sec:metal}
\begin{figure}
\resizebox{\hsize}{!}{\includegraphics{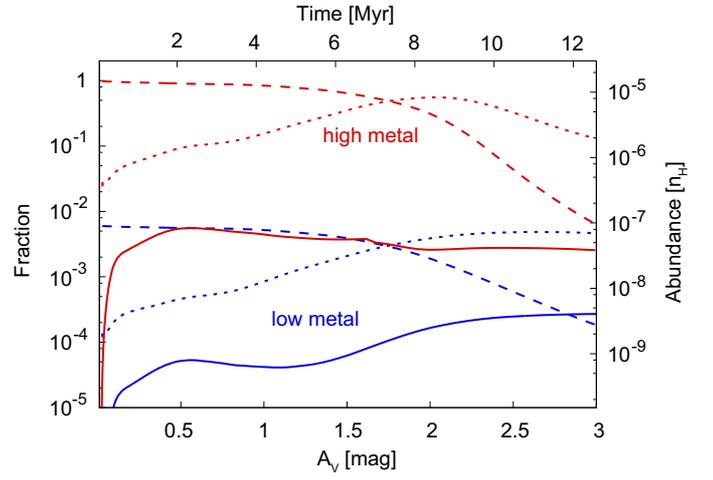}}
\caption{Fraction of elemental sulfur in the form of S$^+$ (dashed lines), neutral atom S (dotted lines), 
and S-bearing molecules (the sum of CS, SO, and H$_2$S; solid lines) in the gas phase as a function of $A_V$ in the fiducial model (blue lines) and model HM (red lines).}
\label{fig:metal}
\end{figure}

\begin{figure}
\resizebox{\hsize}{!}{\includegraphics{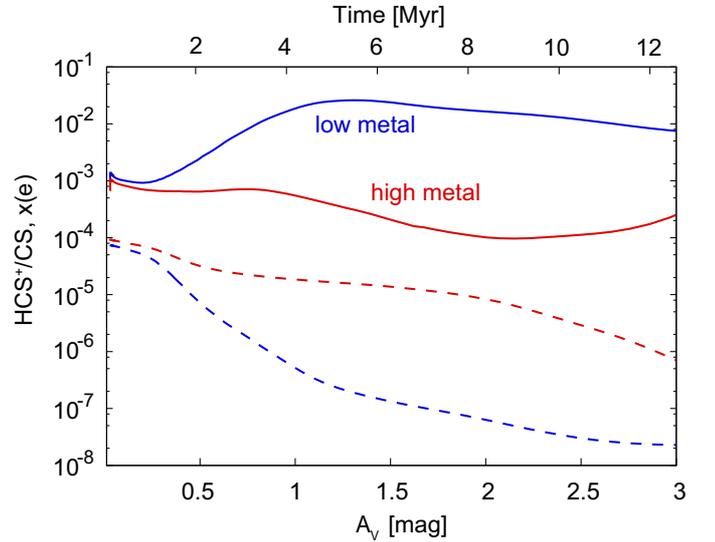}}
\caption{HCS$^+$/CS abundance ratio (solid lines) and electron abundance (dashed lines).
Red lines indicate the model with the high metal abundances, 
while blue lines indicate the fiducial model with the low metal abundances.
See Table \ref{table:element} and Appendix \ref{sec:sulfur}.}
\label{fig:metal2}
\end{figure}

Figure \ref{fig:metal} shows the fraction of elemental sulfur in gaseous species as functions of $A_V$ 
in the model with the high metal abundances (model HM) and in the fiducial model.
The same physical model shown in Figure \ref{fig:phys} is used in model HM. 
Note that in the fiducial model 99.4 \% of elemental sulfur is assumed to be not available for gas and ice chemistry (Table \ref{table:element}).
In model HM sulfur starts to be frozen out onto dust grains after the S$^+$/S transition.
When $A_V$ reaches 3 mag, 85 \% of elemental sulfur is locked in ice predominantly as HS and H$_2$S, 
while 15 \% of sulfur remains in the gas phase.
Then model HM is much richer in sulfur in the gas phase than the fiducial model.
Figure \ref{fig:metal2} compares the abundance of electrons in model HM with that in the fiducial model.
The higher abundance of S$^+$ in model HM keeps the degree of ionization higher than the fiducial model, 
except for the very early stage, where the dominant positive charge carrier is C$^+$.

\begin{figure}
\resizebox{\hsize}{!}{\includegraphics{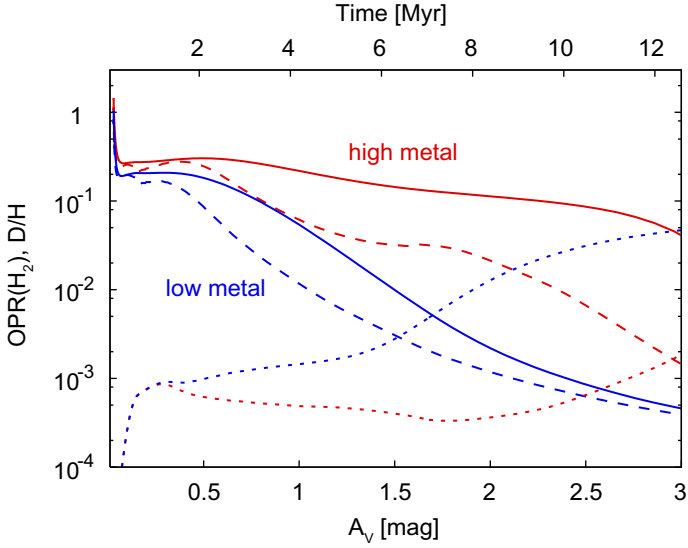}}
\caption{$\op{H_2}$ (solid lines) and atomic D/H ratio (dotted lines) in the gas phase as functions of visual extinction. 
Red lines indicate the high metal case, while blue lines indicate the low metal case. 
The dashed lines indicate the steady-state values of the $\op{H_2}$ given by Equation (\ref{eq:oprsteady}).}
\label{fig:h2opr_hm}
\end{figure}

Figure \ref{fig:h2opr_hm} shows $\op{H_2}$ in model HM as a function of $A_V$.
The $\op{H_2}$ in model HM is much higher at $A_V \gtrsim 0.5$ mag than the fiducial model;
the main destroyers of $\ohh$, H$^+$ and H$_3^+$, are much less abundant in model HM because of the abundant gaseous sulfur.
H$^+$ is efficiently destroyed by the charge transfer to S atoms and to CS, which have lower ionization energies than atomic hydrogen.
The higher degree of ionization leads to the reduced abundance of H$_3^+$ through dissociative electron recombination.
Reflecting the higher $\op{H_2}$, the atomic D/H ratio is lower in model HM than the fiducial model especially at $A_V \gtrsim 1.5$ mag.

Which model yields the better prediction for the $\op{H_2}$ evolution in the ISM?
In Appendix \ref{sec:sulfur}, we compare observations of sulfur-bearing molecules toward diffuse/translucent/molecular clouds 
in the literature with our model results.
We focus on the total abundance of selected sulfur-bearing molecules in the gas phase (CS, SO, and H$_2$S) and the HCS$^+$/CS abundance ratio.
The former can probe the total abundance of elemental sulfur in the gas phase, 
while the latter can probe the ionization degree of the gas, which is related to the S$^+$ abundance.
We find that the fiducial model better reproduces the sulfur observations than model HM, 
and conclude that the fiducial model gives better predictions for the $\op{H_2}$ evolution in the cold ISM.

%\subsubsection{Effect of PAHs}
%{\bf In order to check the effect of PAH chemistry on the conclusion in Section \ref{sec:sulfur},
%we rerun model HM with the PAH chemistry.
%We assume that a single PAH has 30 carbon atoms with the radius of 4 $\AA$ and the abundance of the PAHs is set to be 
%$3\times10^{-7}$ with respect to hydrogen nuclei \citep{draine07,wakelam08}.
%The three charge states of PAHs are considered; neutral (PAH$^0$), negatively charged (PAH$^-$), and positively charged (PAH$^+$) ones.
%We take into account photoionization of PAH$^0$, photodetachment of PAH$^-$, electron attachment of PAH$^0$, electron recombination of PAH$^+$,
%and charge exchange of PAH$^0$ and PAH$^-$ with atomic and molecular ions.
%The reaction rates are calculated based on \citet{bakes98}.
%We run the two extreme cases for the trapping of PAHs in ice mantle;
%all PAHs remains in the gas phase during the simulation or PAH$^0$ is depleted from the gas phase on the adsorption timescale.
%In the latter case, we do not consider any deporption mechanisms for icy PAH$^0$.}

%\citet{albertsson14b} studied temporal evolution of H$_2$ opr in diffuse clouds, using the pseudo-time-dependent model.
%They found that temporal evolution of H$_2$ opr does not significantly depend on the elemental abundances of heavy metals.
%Our result is consistent with theirs; the H$_2$ opr is similar between model HM and the fiducial model at low $A_V \lesssim 0.5$ mag, 
%where deuterium atoms are still one of the main destroyers of H$^+$ and the carbon ion is the dominant charge carrier.

\subsection{The number of active surface ice layers}
The parameter $N_{\rm act}$ was set to be four in the fiducial model.
In the model with $N_{\rm act} = 1$, the timescale in which surface icy molecules are buried in the inactive inert ice mantles is 
shorter than in the fiducial model.
This limits the number of the photodissociation-reformation cycles so as to decrease the HDO/H$_2$O ratio in the active ice layers.
However, the effect on the resultant HDO/H$_2$O ratio in the bulk ice mantle is minor (within a factor of two).

\begin{figure}
\resizebox{\hsize}{!}{\includegraphics{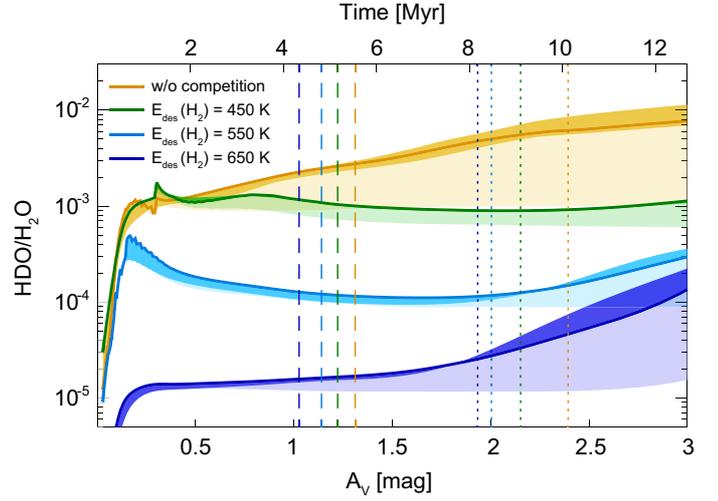}}
\caption{$\hdo$ ratio in the bulk ice in the two grids (with and without the reaction-diffusion competition) of models as functions of $A_V$.
For the case without the competition (yellow), the solid line represents the model with $N_{\rm act} = 4$, $E_{\rm des}({\rm H_2}) = 550$ K, 
and the low-metal abundances.
The areas represent the models including variations in $N_{\rm act}$ and $E_{\rm des}({\rm H_2})$, 
while the light-colored areas represent the models including variations in $N_{\rm act}$, $E_{\rm des}({\rm H_2})$, and the initial abundances.
For the case with the competition, the models with different $E_{\rm des}({\rm H_2})$ are shown by different colors.
The solid lines represent the models with $N_{\rm act} = 4$. 
The areas represent models including the variations in $N_{\rm act}$, 
while the light-colored areas represent the models including the variations in $N_{\rm act}$ and the initial abundances.
The vertical dashed and dotted lines indicate the regions where the fraction of elemental oxygen in the atomic form becomes 
less than 10 \% and 1 \%, respectively, in the models represented by the solid lines.}
\label{fig:waterdh_uncertain}
\end{figure}

\subsection{Desorption energies and reaction-diffusion competition}
\label{sec:edes}
As discussed in Section \ref{sec:dh} and shown in Figure \ref{fig:waterdh_analytic}, 
water ice deuteration depends on the main formation pathway of water ice.
The critical parameter is $\Gamma = R_{\rm OH+H_2}/R_{\rm OH+H}$.
The exact value of $\Gamma$ at given physical conditions is highly uncertain at present; 
it strongly depends on chemical parameters and the treatment of reactions with activation energy barriers.
In this subsection, we explore the impact of desorption energies of H$_2$ and the reaction-diffusion competition on water ice deuteration.
For simplicity, we assume that the desorption energies of atomic and molecular hydrogen are the same with 
a fixed hopping-to-desorption energy ratio of 0.5.
In this case, $\Gamma$ is reduced to be $\varepsilon^{\rm act}_{\rm OH+H_2} \times \pop{H_2}/\pop{H}$, 
where $\pop{A}$ is the population of species A in the active surface ice layers.
Then, a smaller desorption energy leads to a smaller $\Gamma$ (see Figure \ref{fig:rdcomp}).
If the reaction-diffusion competition is turned off, $\Gamma$ becomes smaller.

Figure \ref{fig:waterdh_uncertain} shows the impact of $E^0_{\rm des}({\rm H_2})$ on the $\hdo$ ratio in the bulk ice mantle. 
Let us define the time when the main formation stage of water ice finishes as the time when 
the abundance of atomic oxygen becomes less than 1 \% of the elemental oxygen available for gas and ice chemistry.
This happens at $A_V = 2$ mag - 2.5 mag in our models.
At that time, the $\hdo$ ratios of the bulk ice mantle are $\sim$10$^{-3}$, $\sim$10$^{-4}$, 
and $\sim$10$^{-5}$ in the models with $E^0_{\rm des}({\rm H_2}) = 450$ K, 550 K, and 650 K, respectively.
It is clear that the $\hdo$ ratio is sensitive to $E^0_{\rm des}({\rm H_2})$, while the fraction of elemental oxygen locked 
in water ice is not sensitive to $E^0_{\rm des}({\rm H_2})$. 
The models without the reaction-diffusion competition are also shown in Figure \ref{fig:waterdh_uncertain}.
In this case, we do not see a strong dependence of the $\hdo$ ratio on $E^0_{\rm des}({\rm H_2})$,
since $\Gamma < 1$ regardless of $E^0_{\rm des}({\rm H_2})$ in the range of 450 K - 650 K.
In other words, Reaction (\ref{react:oh+h}) dominates over Reaction (\ref{react:oh+h2}), 
so that the $\hdo$ ratio is more easily enhanced by a high atomic D/H ratio (see Figure \ref{fig:waterdh_analytic}).
When the water ice formation is finished, the $\hdo$ ratio in the bulk ice mantle is $6\times10^{-3}$, which is
higher than the models with the reaction-diffusion competition.
 
The observationally derived upper limits of the $\hdo$ ice ratio in the cold outer envelope 
of embedded protostars are $(2-10)\times10^{-3}$ \citep[e.g.,][]{dartois03},
while there is still no observational constraint for the $\hdo$ ice ratio in molecular clouds.
Let us assume that the $\hdo$ ice ratio in molecular clouds is less than $2\times10^{-3}$.
In Figure \ref{fig:waterdh_uncertain}, we can see that this low $\hdo$ ice ratio is reproduced by models 
with the reaction-diffusion competition, in which Reaction (\ref{react:oh+h2}) contribute significantly to the water formation, 
while the ratio is reproduced only in a limited parameter space in the models without competition.

In addition, the present work suggests the possibility that water ice formed in molecular clouds is not so enriched in deuterium, 
compared to the water vapor observed in the vicinity of low-mass protostars where water ice is sublimated \citep[$\sim$10$^{-3}$, e.g.,][]{persson13}.
If this is the case, the enrichment of deuterium in water ice should mostly occur in the later prestellar core or/and protostellar phases, 
where interstellar UV radiation is heavily attenuated, CO is frozen out, and the $\op{H_2}$ is lower than in molecular clouds.
Note that even if water ice formed in molecular clouds is not enriched in deuterium compared to the elemental D/H ratio in the local ISM ($1.5\times10^{-5}$), 
an additional formation on small amount of water ice with a high $\hdo$ ratio in prestellar/protostellar core, 
e.g., $x({\rm H_2O\,ice})=5\times10^{-6}$ with $\hdo = 2\times10^{-2}$, 
can explain the observationally derived $\hdo$ ratio in the vicinity of the low-mass protostars (Furuya et al. in prep).
%In addition, the effect of the different physical parameters, such as the velocity of accretion flow, radiation field, and cosmic-ray ionization rate, 
%on the $\hdo$ ratio and H$_2$ opr should be explored in future work.

\subsection{Ram pressure due to the accretion flow}
\label{sec:weaker_shock} 
In our fiducial physical model, the density of pre-shock materials and the velocity of accretion flow are 
assumed to be $n_0 = 10$ cm$^{-3}$ and $v_0 = 15$ km s$^{-1}$, respectively. 
The resultant post-shock material has the relatively high density of $\sim$10$^4$ cm$^{-3}$, compared to 
the average density of giant molecular clouds derived from observations \citep[][and references therein]{bergin07}. 
In this subsection, we briefly present a model with a smaller ram pressure, $n_0 = 3$ cm$^{-3}$ and $v_0 = 10$ km s$^{-1}$ (model 3 in H10),
in which the density of post-shock materials is $\sim$10$^3$ cm$^{-3}$.
The goal is to check if our primary results presented in Section \ref{sec:result} do not change qualitatively.
In this model, the internal pressure due to self gravity overwhelms the ram pressure at $A_V > 1.6$ mag \citep[cf.][]{hartmann01}.
Nevertheless, we use the shock model until $A_V$ reaches 3 mag.
 
\begin{figure*}
%\centering
\sidecaption
\resizebox{\hsize}{12cm}{\includegraphics{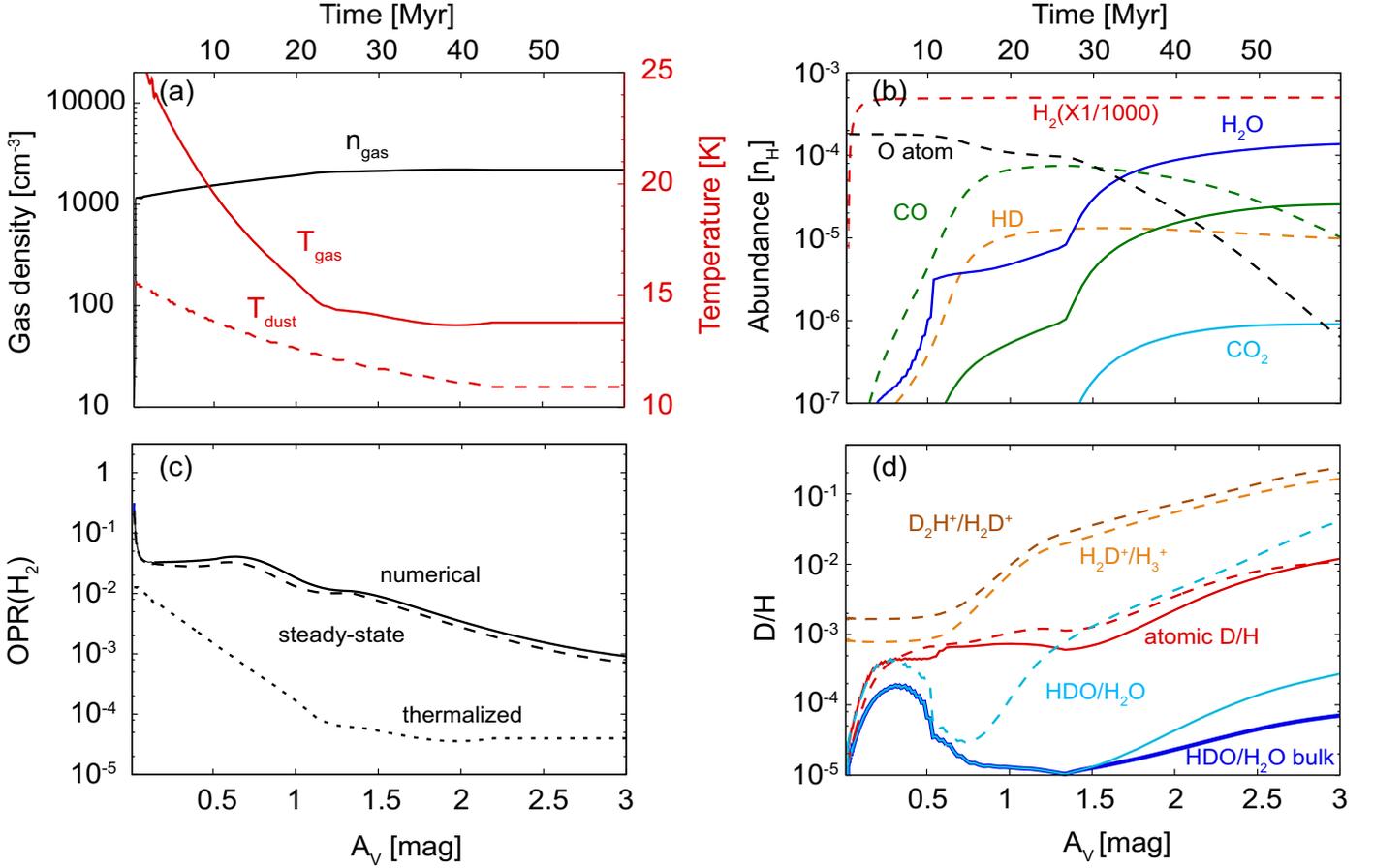}}
\caption{Results from a weaker shock model plotted vs visual extinction to be compared with analogous figures for our fiducial model. 
(a) Physical evolution after passing through the shock front (see Figure \ref{fig:phys}).
(b) Fractional abundances of selected species with respect to hydrogen nuclei (see Figure \ref{fig:ab}).
(c) The $\op{H_2}$ (see Figure \ref{fig:opr}).
(d) Deuteration ratios of selected species (see Figure \ref{fig:dh}).}
\label{fig:weaker_shock}
\end{figure*}

Figure \ref{fig:weaker_shock} summarizes the results from the model with the smaller ram pressure.
We confirmed that most of our qualitative chemical results presented earlier in this paper do not change. 
The main differences are 
(i) molecular formation requires a larger shielding column density because of the lower gas density (B04; H10),
(ii) the $\op{H_2}$ is lower when most hydrogen is locked in H$_2$ reflecting higher $\xi_{\rm H_2}/n_{\rm H}$ 
(Equation (\ref{eq:opr_atomic})), and
(iii) the $\op{H_2}$ is almost at steady-state during the simulation because of the longer timescale of the physical evolution.

\section{Comparisons with previous theoretical studies}
\label{sec:comparison}
Water ice deuteration has been numerically studied over wide ranges of the physical conditions: 
$n_{\rm H}$ = $10^4$-$10^6$ cm$^{-3}$, $T$ = 10-20 K, and $A_V$ = a few-10 mag.
Most of the previous studies used a pseudo-time-dependent chemical model, in which physical conditions are constant 
with time \citep[e.g.,][]{taquet13b,sipila13,lee15}, 
while some investigated deuterium chemistry in gravitationally collapsing clouds/cores \citep[e.g.,][]{aikawa12b,taquet14}.
The general conclusions of the previous studies are that (i) the $\hdo$ ice ratio is controlled 
by the temporal evolution of the atomic D/H ratio \citep[one exception is][but see below]{cazaux11}
and 
(ii) the evolution of the atomic D/H ratio strongly depends on the initial $\op{H_2}$, which is treated as a free parameter.
The rationale for point (i) is that H$_2$O and HDO are predominantly formed via sequential reactions of 
atomic hydrogen/deuterium with atomic oxygen in the previous models.
 
The present study has considered the chemistry during which cold and dense conditions are first achieved by the passage
of a shock through precursor \mbox{\ion{H}{i}}-dominated clouds.
This approach allows us to study the evolution of $\op{H_2}$ and water ice deuteration without arbitrary assumptions concerning the initial $\op{H_2}$.
Contrary to point (i), the present study has shown that the $\hdo$ ice ratio is not always controlled by the atomic D/H ratio.
We have shown that deuterium can be removed from water ice chemistry on the timescale of water ice photodissociation,
when Reaction (\ref{react:oh+h2}) dominates over Reaction (\ref{react:oh+h}) or equivalently $\Gamma > 1$.
As a consequence, the $\hdo$ ratio in the active surface ice layers can be much smaller than the atomic D/H ratio.
On the other hand, when $\Gamma < 1$, the $\hdo$ ice ratio scales with the atomic D/H ratio as predicted by the previous studies.
The former condition ($\Gamma > 1$) favors gas with high but not very high extinction, where the dust temperature is low and the gaseous H/H$_2$ abundance ratio is low, 
while the latter condition favors gas with a low extinction.
The onset of water ice mantle formation requires a threshold extinction, which depends on the UV field, the gas density, 
and the dust temperature \citep[e.g.,][]{tielens05}.
Therefore, the critical question here is what fraction of water ice is formed with $\Gamma > 1$ in molecular clouds.
In our fiducial case, for example, the majority of water ice is formed with $\Gamma > 1$.
However, the quantification depends on uncertain chemical parameters, 
such as the hopping timescale of H and H$_2$ on a surface and the tunneling transmission probability of Reaction (\ref{react:oh+h2}), 
as well as adopted physical models.
We also note that when the reaction-diffusion competition is turned off, $\Gamma$ is always less than unity in our models.
To the best of our knowledge, no previous study focusing on ice deuteration considered the reaction-diffusion competition \citep[except for][]{cazaux11}.
This explains why the previous studies commonly claimed point (i).

\citet{cazaux11} studied the deuteration of water ice in a translucent cloud followed by gravitational collapse.
They claimed that the $\hdo$ bulk ice ratio scales with the D/H$_2$ ratio at $T_{\rm dust} \lesssim 15$ K, 
while it scales with the atomic D/H ratio at higher dust temperatures.
They demonstrated that for $T_{\rm dust} \lesssim 15$ K, OH and OD are mainly formed by 
$\react{O}{H_2}{OH}{H}$ and O + D $\rightarrow$ OD, respectively, followed by hydrogenation to form H$_2$O and HDO.
At higher dust temperatures, OH and OD are mainly formed by the reaction of atomic hydrogen (deuterium) with atomic oxygen in their model.
At $T_{\rm dust} \lesssim 15$ K, the $\hdo$ bulk ice ratio is $\sim$10$^{-6}$ at the end of their simulations, 
while it is $\sim$10$^{-3}$ at $T_{\rm dust} \gtrsim 15$ K.
The key assumption they made was that the surface reaction, $\react{O}{H_2}{OH}{H}$, occurs efficiently through quantum tunneling.
Since the reaction $\react{O}{H_2}{OH}{H}$ is endothermic by 960 K, 
the validity of the assumption has been questioned by several authors \citep{oba12,taquet13a,lamberts14}.
Recently, \citet{lamberts14} concluded that the contribution of the O + H$_2$ pathway to the total OH formation in the ISM is small (likely much less than 10 \%),
based on their experiments and modeling.

\section{Summary}
\label{sec:conclusion}
We have investigated water deuteration and the $\op{H_2}$ during the formation and early evolution of a molecular cloud, 
following the scenario that \mbox{\ion{H}{i}}-dominated gas is swept and accumulated by global accretion flows to form molecular clouds.
We used the one-dimensional shock model developed by \citet{bergin04} and \citet{hassel10} to follow the physical evolution
of post-shock materials, combined with post-processing detailed gas-ice chemical simulations.
Our findings are summarized as follows.

\begin{enumerate}
{\item In the H/H$_2$ transition region, the $\op{H_2}$ is already much less than the statistical value of three, 
because the timescale of spin conversion through the reaction of $\ohh$ with H$^+$ is shorter than the timescale of
H$_2$ formation on the grain surface.
When most hydrogen is locked in H$_2$, the $\op{H_2}$ is nearly at steady state with the value of $\sim$0.1 in our fiducial model.
The exact value depends on the ionization conditions and the gas temperature (see Appendix \ref{appendix:h2_opr_transition} and Figure \ref{fig:h2opr_steady}).
At later times, the $\op{H_2}$ decreases in a non-equilibrium manner as the gas accumulates, reflecting local physical and chemical conditions.
%It reaches 10$^{-3}$ when the visual extinction of the post-shock materials reaches 3 mag ($t = 10$ Myr after passing through the shock front in our fiducial model).
The evolution of the $\op{H_2}$ depends on the abundances of elements heavier than oxygen, especially sulfur.
A higher sulfur abundance leads to the decrease of the abundances of H$^+$ and H$_3^+$, which leads to a longer ortho-para spin conversion timescale in the gas phase.
}
{\item The $\hdo$ ratio in the bulk ice is as low as 10$^{-4}$ at the end of our fiducial simulation where most of oxygen is already locked up in molecules.
The key mechanism to suppress water ice deuteration is the cycle of photodissociation and reformation of water ice on the icy surface, 
which removes deuterium from water ice chemistry at the timescale of photodissociation.
The efficiency of the mechanism depends on which formation path, the barrierless reaction OH + H $\rightarrow$ H$_2$O 
or the barrier-mediated reaction OH + H$_2$ $\rightarrow$ H$_2$O + H, is more effective,
because the pathway including barrier-mediated reactions favors hydrogenation over deuteration. 
Depending on the contribution of the OH + H$_2$ pathway to water ice formation, 
the resultant $\hdo$ ratio in the bulk ice ranges from 10$^{-5}$ to 10$^{-3}$.
The contribution of the OH + H$_2$ pathway to water ice formation depends strongly on 
whether or not we include the reaction-diffusion competition and hopping timescale of H$_2$ on the surface.
The $\op{H_2}$ plays a minor role in water ice deuteration, because the production of atomic deuterium is dominated by photodissociation of HD 
at the main formation stage of water ice, rather than the electron dissociative recombination of deuterated ions.
}
{\item The above results suggest the possibility that water ice formed in molecular clouds is deuterium-poor, 
compared to the water vapor observed in the vicinity of protostars where water ice is sublimated.
If this is the case, the enrichment of deuterium in water should mostly occur at somewhat later evolutionary stages of 
star formation, i.e., cold prestellar/protostellar cores, 
where interstellar UV radiation is heavily attenuated, CO is frozen out, and $\op{H_2}$ is lower 
than in molecular clouds.
When the protostellar core begins to warm up, the situation becomes more complex.
}
{\item The $\hdo$ ratio in the gas phase, the active surface ice layers, and in the bulk ice mantle can have different values. 
Even if the H$_2$O vapor abundance is determined by the balance between photodesorption and photodissociation,
it does not necessarily mean that the $\hdo$ ratio in the gas phase and in the surface ice layers are similar.
}
\end{enumerate}
 
\begin{acknowledgements}
K.F. thanks Ewine F. van Dishoeck, Catherine Walsh, and Maria N. Drozdovskaya for stimulating discussions, 
Carina Arasa for providing useful data on photolysis, and Vianney Taquet for constructive comments on the manuscript.
We thank the anonymous referee for helpful suggestions on the manuscript.
K.F. is supported by the Research Fellowship from the Japan Society for the Promotion of Science (JSPS).
Y.A. acknowledges the support by JSPS KAKENHI Grant Numbers 23103004 and 23540266.
A.V. acknowledges the financial support of the European Research Council (ERC; project PALs 320620).
E.H. acknowledges the support of the National Science Foundation for his program in astrochemistry, 
and support from the NASA Exobiology and Evolutionary Biology program under subcontract from Rensselaer Polytechnic Institute.
Some kinetic data we used have been downloaded from the online database KIDA \citep[][http://kida.obs.u-bordeaux1.fr]{wakelam12}
\end{acknowledgements}

\begin{appendix}
\section{The condition for the nuclear spin ratio of H$_2$ to slow down deuteration processes}
\label{appendix:opr_dh}
Here we derive the condition determining how much $\ohh$ is needed to slow down deuterium fractionation driven by Reaction (\ref{eq:dfrac_reaction}).
It is well established that the destruction of H$_2$D$^+$ by electrons and CO suppresses the fractionation as well as by H$_2$.
H$_2$D$^+$ also reacts with HD to form the doubly deuterated species D$_2$H$^+$, leading to further fractionation.
Considering the balance of formation and destruction, the steady-state H$_2$D$^+$/H$_3^+$ ratio can be calculated as
\begin{align}
\begin{split}
&\frac{\num{\ohhdp} + \num{\phhdp}}{\num{\ohhhp} + \num{\phhhp}} = \\
&\frac{k_{\ref{eq:dfrac_reaction}f}\num{HD}}{k_{\rm CO}\num{CO}+k_{\rm e}\num{e}+k_{\rm HD}\num{HD}+k_{\ref{eq:dfrac_reaction}b}\num{H_2}}, \label{eq:dh_steady}
\end{split}
\end{align}
where $\num{i}$ is the number density of species i, $k_{\rm CO}$ is the rate coefficient of H$_2$D$^+$ with CO, 
$k_{\rm e}$ is the rate coefficient of dissociative electron recombination of H$_2$D$^+$,
$k_{\rm HD}$ is the rate coefficient of H$_2$D$^+$ with HD,
and $k_{\ref{eq:dfrac_reaction}f}$ and $k_{\ref{eq:dfrac_reaction}b}$ are the effective rate coefficients 
for Reaction (\ref{eq:dfrac_reaction}) in the forward direction and in the backward direction, respectively.
Since the forward Reaction (\ref{eq:dfrac_reaction}) is exothermic, $k_{\ref{eq:dfrac_reaction}f}$ does not depend on the $\op{H_2}$.
We can say that the presence of $\ohh$ slows down the fractionation when $k_{\ref{eq:dfrac_reaction}b}\num{H_2}$ is greater than 
$k_{\rm CO}\num{CO}$, $k_{\rm e}\num{e}$, and $k_{\rm HD}\num{HD}$.

The endothermicity of Reaction (\ref{eq:dfrac_reaction}) in the backward direction depends on the nuclear spin states of H$_2$ and H$_2$D$^+$.
The following set of reactions can occur:
\begin{align}
\reactalign{\phhdp}{\phh}{\phhhp}{HD} - 232 \,\,{\rm K},\label{react:back_pp} \\ 
\reactalign{\phhdp}{\ohh}{\ohhhp \,\, (\phhhp)}{HD} - 95 \,\, (-62) \,\,{\rm K},\label{react:back_po} \\ 
\reactalign{\ohhdp}{\phh}{\ohhhp \,\, (\phhhp)}{HD} - 178 \,\, (-145) \,\,{\rm K},\label{react:back_op}\\
\reactalign{\ohhdp}{\ohh}{\ohhhp \,\, (\phhhp)}{HD} - 8 \,\, (+25) \,\,{\rm K}.\label{react:back_oo}
\end{align}
The rate coefficients of the reactions can be found in \citet{hugo09}, who calculated them with the assumption that the reaction proceeds 
by a scrambling mechanism in which all protons are equivalent; 
results are different if long-range hopping is the effective mechanism.
The effective rate coefficient $k_{\ref{eq:dfrac_reaction}b}$ can be defined as follows \citep{gerlich02,lee15}:
\begin{align}
\begin{split}
k_{\ref{eq:dfrac_reaction}b}\num{H_2D^+}\num{H_2} = & k_{\rm \ref{react:back_pp}}\num{\phhdp}\num{\phh} \\
                                                    & + k_{\rm \ref{react:back_po}}\num{\phhdp}\num{\ohh} \\
                                                    & + k_{\rm \ref{react:back_op}}\num{\ohhdp}\num{\phh} \\
                                                    & + k_{\rm \ref{react:back_oo}}\num{\ohhdp}\num{\ohh} \label{eq:kb_def}.
\end{split}
\end{align}
From Equation (\ref{eq:kb_def}), we can express $k_{\ref{eq:dfrac_reaction}b}$ as a function of ortho-to-para ratios of H$_2$ and H$_2$D$^+$ ($\op{H_2D^+}$), 
and gas temperature.
%\begin{align}
%k_{\ref{eq:dfrac_reaction}b} = \frac{k_{\rm \ref{react:back_pp}} + \op{H_2} k_{\rm \ref{react:back_po}} + \op{H_2D^+} k_{\rm \ref{react:back_op}} 
%+ \op{H_2}\op{H_2D^+} k_{\rm \ref{react:back_oo}}}{(1+\op{H_2})(1+\op{H_2D^+})}. \label{eq:kb}
%\end{align}

The $\op{H_2D^+}$ in the dense ISM is mainly determined by the following reactions \citep{gerlich02,hugo09}:
\begin{align}
\reactalign{\phhdp}{\phh}{\ohhdp}{\ohh} - 257 \,\,{\rm K},\label{react:h2dp_pp} \\
\reactalign{\phhdp}{\ohh}{\ohhdp}{\ohh \,\, (\phh)} - 87 \,\, (+83) \,\,{\rm K},\label{react:h2dp_po} \\ 
\reactalign{\ohhdp}{\phh}{\phhdp}{\ohh} - 83 \,\,{\rm K} ,\label{react:h2dp_op} \\
\reactalign{\ohhdp}{\ohh}{\phhdp}{\phh \,\, (\ohh)} +257 \,\, (+87) \,\,{\rm K}. \label{react:h2dp_oo}
\end{align} 
Then, assuming steady state, the $\op{H_2D^+}$ is given as a function of $\op{H_2}$ and gas temperature:
\begin{align}
\opst{H_2D^+} \approx \frac{ k_{\rm \ref{react:h2dp_pp}} + \op{H_2} k_{\rm \ref{react:h2dp_po}}}
{k_{\rm \ref{react:h2dp_op}} + \op{H_2} k_{\rm \ref{react:h2dp_oo}}}. \label{eq:op_h2dp}
\end{align}
From Equations (\ref{eq:kb_def}) and (\ref{eq:op_h2dp}), we can express $k_{\ref{eq:dfrac_reaction}b}$ 
as a function of $\op{H_2}$ and the gas temperature.
In Figure \ref{fig:critical_opr}, we show threshold abundances of CO and electrons as functions of the $\op{H_2}$, 
below which Reaction (\ref{eq:dfrac_reaction}) in the backward reaction is more efficient than the destruction by CO and electrons.
Roughly speaking, the backward reaction rate exceeds the destruction rate by CO when $\op{H_2} \gtrsim 20\num{CO}/\num{H_2}$ at $\le$20 K, 
while it exceeds the recombination rate with electrons when $\op{H_2} \gtrsim 3 \times 10^{3}\num{e}/\num{H_2}$ at $\le$20 K.
Assuming the canonical HD abundance with respect to H$_2$, $3\times 10^{-5}$, 
the condition $k_{\ref{eq:dfrac_reaction}b}\num{H_2} > k_{\rm HD}\num{HD}$ corresponds to $\op{H_2} \gtrsim 6\times 10^{-4}$ at $\le$20 K.

\section{Analytical treatment of H$_2$ ortho-to-para ratio}
\label{appendix:h2_opr}
In this appendix, we derive an analytical solution of $\op{H_2}$ when abundances of ionic species and the H$_2$ formation rate on grain surfaces are given.
Let us consider the following set of differential equations, which describe the temporal variations 
of the abundances of $\ohh$ and $\phh$ in the gas and solid phases:
\begin{align}
%\frac{d\num{\phh}}{dt} &= \alpha^{\rm eva}_{\phh} - \alpha^{\rm acc}_{\phh} + \sum_i k_i^{\rm o\rightarrow p} n_i \num{\ohh} 
%- \sum_j k_j^{\rm p\rightarrow o} n_j \num{\phh} - \sum_l k_l^{\rm des}\num{\phh},\\
%\frac{d\num{\ohh}}{dt} &= \alpha^{\rm eva}_{\ohh} - \alpha^{\rm acc}_{\ohh} - \sum_i k_i^{\rm o\rightarrow p} n_i \num{\ohh} 
%+ \sum_j k_j^{\rm p\rightarrow o} n_j \num{\phh} - \sum_l k_l^{\rm des}\num{\ohh},\\
\ddt \num{\ohh} =& W({\ohh}) - F({\ohh}) + \tau^{-1}_{\rm p\rightarrow o}\num{\phh} \nonumber \\
                 &- \tau^{-1}_{\rm o\rightarrow p}\num{\ohh} - D({\ohh}), \label{eq:ap1}\\
\ddt \num{\phh} =& W({\phh}) - F({\phh}) + \tau^{-1}_{\rm o\rightarrow p}\num{\ohh} \nonumber \\
                 &- \tau^{-1}_{\rm p\rightarrow o}\num{\phh} - D({\phh}), \label{eq:ap2}\\
\ddt (\pop{\ohh} n_{\rm gr}) =&  F({\ohh}) - W({\ohh}) + b_{\rm o} R_{\rm H_2}, \label{eq:ap3}\\
\ddt (\pop{\phh} n_{\rm gr}) =&  F({\phh}) - W({\phh}) + (1-b_{\rm o}) R_{\rm H_2}, \label{eq:ap4}
\end{align}
where $F$, $W$, $D$ are the adsorption rate, desorption rate, and rate of H$_2$ destruction via e.g., photodissociation, respectively. 
$\langle A \rangle$ in Equations (\ref{eq:ap3}) and (\ref{eq:ap4}) is the population of species A in the active surface ice layers, 
and thus $\langle A \rangle n_{\rm gr}$ is the number density of species A in the surface ice layers per unit volume of gas.
$R_{\rm H_2}$ is the formation rate of H$_2$ on grain surfaces, while $b_{\rm o}$ is the branching ratio to form $\ohh$.
In Equations (\ref{eq:ap1}) and (\ref{eq:ap2}), we neglect H$_2$ formation in the gas phase for simplicity. 
In Equations (\ref{eq:ap3}) and (\ref{eq:ap4}), we do not consider nuclear spin conversion on the surface, 
though it is straightforward to include the spin conversion on the surface in the following analysis.

Combining Equations (\ref{eq:ap1})-(\ref{eq:ap4}), we get 
\begin{equation}
\begin{split}
\ddt \op{H_2} =& \frac{d}{dt}\left(\frac{\num{\ohh}}{\num{\phh}}\right) \\
       \approx & \frac{d}{dt}\left(\frac{\num{\ohh} + \pop{\ohh}n_{\rm gr}}{\num{\phh} + \pop{\phh}n_{\rm gr}}\right) \\
       \approx & \frac{1}{\num{\phh}}\frac{d}{dt}\left(\num{\ohh} + \pop{\ohh}n_{\rm gr} \right) \\
               & - \frac{\num{\ohh}}{(\num{\phh})^2}\frac{d}{dt}\left(\num{\phh} + \pop{\phh}n_{\rm gr}\right) \\
%& = -\left(\sum_i k_i^{\rm o\rightarrow p}n_i + (1-b_r)\frac{R^{\rm form}_{\rm H_2}}{\num{H_2}}\right) r_{\rm o-p}^2 \\
%    & -\left(\sum_i k_i^{\rm o\rightarrow p}n_i + \sum_j k_j^{\rm p\rightarrow o}n_j + (2b_r - 1)\frac{R^{\rm form}_{\rm H_2}}{\num{H_2}}\right) r_{\rm o-p} \\
%    & + \sum_j k_j^{\rm p\rightarrow o}n_j + b_rR^{\rm form}_{\rm H_2}
              =& -[\tau^{-1}_{\rm o\rightarrow p} + (1-b_{\rm o})\tau^{-1}_{\rm H_2}] (\op{H_2})^2 \\
               & -[ \tau^{-1}_{\rm o\rightarrow p} - \tau^{-1}_{\rm p\rightarrow o} - (2b_{\rm o} - 1)\tau^{-1}_{\rm H_2}] \op{H_2} \\
               & + [\tau^{-1}_{\rm p\rightarrow o} + b_{\rm o}\tau^{-1}_{\rm H_2}], \label{eq:dopdt}
\end{split}
\end{equation}
where we used $\num{\ohh} + \num{\phh} =\num{H_2}$ and $\num{H_2} \gg \pop{H_2} n_{\rm gr}$.
The latter should be valid in molecular gases; the binding energy of H$_2$ on a H$_2$ substrate is only 23 K, 
corresponding to an sublimation timescale of $\sim$10$^{-10}$ s at $T = 10$ K \citep{vidali91,cuppen07},
while the adsorption timescale is $\sim$10$^9/n_{\rm H}$ yr.
The H$_2$ formation timescale, $\tau_{\rm H_2}$, was defined as $\num{H_2}/R_{\rm H_2}$.
We also assumed that the rate coefficients of reactions to destroy $\ohh$ and $\phh$ are the same, i.e., $D(\ohh)/D(\phh) = \num{\ohh}/\num{\phh}$.
Equation (\ref{eq:dopdt}) describes the time evolution of the $\op{H_2}$.
The terms $\tau_{\rm o\rightarrow p}$, $\tau_{\rm p\rightarrow o}$, and $\tau_{\rm H_2}$ are time-dependent in general.
It is straightforward to solve Equation (\ref{eq:dopdt}) in the astrochemical simulations without nuclear spin state chemistry,
and one may obtain a reasonable approximation of the temporal variation of the $\op{H_2}$.
In order to solve Equation (\ref{eq:dopdt}) analytically, we consider $\tau_{\rm o\rightarrow p}$, $\tau_{\rm p\rightarrow o}$, and $\tau_{\rm H_2}$ as constant.
This assumption is valid when the ortho-to-para spin conversion time scale is longer than 
the chemical (formation and destruction) timescale of hydrogen and light ions.
The solution is given as follows:
\begin{align}
\op{H_2}(t) &= \frac{\opst{H_2} + \alpha \exp(-t/\tau_{\rm opr})}{1 - \alpha \exp(-t/\tau_{\rm opr})}, \label{eq:oprtevol} \\
\frac{1}{\tau_{\rm opr}} &= \frac{1}{\tau_{\rm p\rightarrow o}} + \frac{1}{\tau_{\rm o\rightarrow p}} + \frac{1}{\tau_{\rm H_2}}, \\
%\tau_{\rm o\rightarrow p} &= \frac{\num{\ohh}}{R_{\rm o\rightarrow p}}, \,\, \tau_{\rm p\rightarrow o} = \frac{\num{\phh}}{R_{\rm p\rightarrow o}}, \,\,
%\tau_{\rm H_2} = \frac{\num{H_2}}{R_{\rm H_2}},\\
%\alpha &\equiv \op{H_2}(t \to \infty) = \frac{\tau^{-1}_{\rm p\rightarrow o} 
%+ b_{\rm o} \tau^{-1}_{\rm H_2}}{\tau^{-1}_{\rm o\rightarrow p} + (1-b_{\rm o})\tau^{-1}_{\rm H_2}}, \label{eq:oprsteady}\\
%\alpha &\equiv \op{H_2}(t \to \infty) = \frac{9\exp(-170/T)+ b_{\rm o}\tau_{\rm o\rightarrow p}/\tau_{\rm H_2}}
%{1 + (1-b_{\rm o})\tau_{\rm o\rightarrow p}/\tau_{\rm H_2}}, \label{eq:oprsteady}\\
\alpha &= \frac{\opini{H_2}-\opst{H_2}}{\opini{H_2}+1},
\end{align}
where $\tau_{\rm opr}$ gives the characteristic timescale of $\op{H_2}$ evolution, 
and $\opini{H_2}$ is the initial $\op{H_2}$.
The steady-state value of $\op{H_2}$ ($\opst{H_2} \equiv \op{H_2}(t \to \infty)$) is given in Equation (\ref{eq:oprsteady}), 
which was derived by \citet{lebourlot91} in a different manner.

When $\opst{H_2} \ll \opini{H_2} \ll 1$, Equation (\ref{eq:oprtevol}) can be simply rewritten as $\op{H_2}(t) \approx \opst{H_2} + \opini{H_2}\exp(-t/\tau_{\rm opr}).$
Then, it takes a greater time than $\ln[\opini{H_2}/\opst{H_2}]\tau_{\rm opr}$ to reach the steady state value.

\section{H$_2$ ortho-to-para ratio when the conversion of hydrogen into H$_2$ is almost complete \label{appendix:h2_opr_transition}}
Evolution of molecular abundances have often been investigated via pseudo-time dependent models in which 
hydrogen is assumed to be in H$_2$ at the beginning of the calculation. 
In such models, the molecular D/H ratio depends on the initial $\op{H_2}$, which is treated as a free parameter \citep[e.g.,][]{flower06}.
In this appendix we analytically derive the $\op{H_2}$ when the conversion of hydrogen into H$_2$ is almost complete.

\begin{figure}
\resizebox{\hsize}{!}{\includegraphics{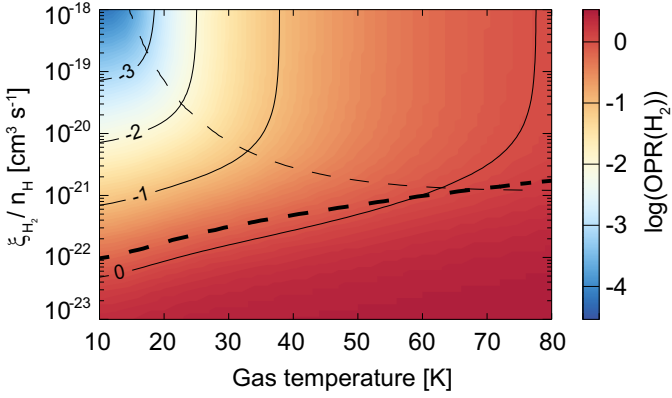}}
\caption{$\op{H_2}$ when the conversion of hydrogen into H$_2$ is almost complete under the steady-state assumption. 
Above the thick dashed line, where $\tau_{\rm o\rightarrow p}$ is smaller than $\tau_{\rm H_2}$, 
the steady-state assumption is justified.
Above the thin dashed line, where $\beta_1$ is larger than $\beta_2$, 
the $\op{H_2}$ is similar to the low-temperature thermalized value of $9\exp(-170.5/T_{\rm gas})$.
See Appendix \ref{appendix:h2_opr_transition}.}
\label{fig:h2opr_steady}
\end{figure}

During the H/H$_2$ transition, ortho-para spin conversion occurs through Reaction (\ref{react:h2+h+}).
H$^+$ is primarily formed via cosmic-ray/X-ray ionization of atomic hydrogen, 
while it is mainly destroyed via recombination with electrons, 
and charge transfer to other species, such as atomic deuterium and oxygen \citep[e.g.,][]{dalgarno73}.
The former is valid when $\num{H}/\num{H_2} > \xi_{\rm H}/(b_{\rm H^+}\xi_{\rm H_2}) \approx 0.05$, 
where $\xi_{\rm H}$ and $\xi_{\rm H_2}$ are the ionization rates of atomic hydrogen and H$_2$, respectively.
$b_{\rm H^+}$ is the branching ratio to form H$^+$ for the ionization of H$_2$.
At steady state, the number density of H$^+$ can be given as
\begin{align}
\num{H^+} &\approx \frac{\xi_{\rm H}\num{H}}{k_{\rm (e+H^+)}\num{\rm C^+} + k_{\rm (D+H^+)}\num{D} + k_{\rm (O+H^+)}\num{O}}, \label{eq:hp}
\end{align}
where
\begin{align}
\begin{split}
\xi_{\rm H}\num{H} =& k_{\rm (e+H^+)}\num{e}\num{H^+} + k_{\rm (D+H^+)}\num{D}\num{H^+} \\
                    &+ k_{\rm (O+H^+)}\num{O}\num{H^+},
\end{split}
\end{align}
where $k_{(X+Y)}$ is the rate coefficient of reaction X + Y.
We assumed that the number density of electrons is equal to that of carbon ions.
The carbon ion is the dominant form of carbon in the H/H$_2$ transition region, while oxygen and deuterium are predominantly in atomic form.
With Equation (\ref{eq:hp}), we can evaluate $\beta_1$ and $\beta_2$ (Equations (\ref{eq:beta1}) and (\ref{eq:beta2})) by the following:
\begin{align}
\beta_1 & = k_{\ref{react:h2+h+}b}/k_{\ref{react:h2+h+}f} = 9\exp(-170.5/T_{\rm gas}), \label{eq:b1_h+}\\
\begin{split}
\beta_2 & = \frac{0.5\num{H} v_{\rm th} \pi a^2 \num{gr} }{k_{\ref{react:h2+h+}f}\num{H^+}\num{H_2}} \\
& \approx \frac{v_{\rm th} \pi a^2 x_{\rm gr}[k_{\rm (e+H^+)}\ab{C^+} + k_{\rm (D+H^+)}\ab{D} + k_{\rm (O+H^+)}\ab{O}]}
{k_{\ref{react:h2+h+}f}\ab{H_2}[\xi_{\rm H_2}/n_{\rm H}]}, \label{eq:b2_h+}
\end{split}
\end{align}
where $\ab{i}$ is the abundance of species i with respect to hydrogen nuclei, 
$n_{\rm H}$ is the number density of hydrogen nuclei, 
$v_{\rm th}$ is the thermal velocity of atomic hydrogen, 
and $k_{\ref{react:h2+h+}f}$ and $k_{\ref{react:h2+h+}b}$ are the rate coefficients of Reaction (\ref{react:h2+h+}) 
in the forward direction and the backward direction, respectively.
We used $\xi_{\rm H_2} = 2.2\xi_{\rm H}$.
We assumed that the H$_2$ formation rate is half of the accretion rate of atomic hydrogen onto dust grains.
By substituting Equations (\ref{eq:b1_h+}) and (\ref{eq:b2_h+}) into Equation (\ref{eq:oprsteady}), 
we can get the $\op{H_2}$ in the H/H$_2$ transition regime under the steady state assumption 
as a function of $\xi_{\rm H_2}/n_{\rm H}$ and gas temperature.
In Figure \ref{fig:h2opr_steady}, we show the $\op{H_2}$ when the the conversion of hydrogen into H$_2$ is 
(almost) complete, i.e., $\ab{H_2}=0.5$.
Above the thick dashed line, where $\tau_{\rm o\rightarrow p}$ is smaller than $\tau_{\rm H_2}$, 
the steady-state assumption is justified.

It is clear that the $\op{H_2}$ is greater than unity only when H$_2$ formation occurs at low ionization 
($\xi_{\rm H_2}/n_{\rm H} < 10^{-22}$ cm$^3$ s$^{-1}$) and/or 
warm conditions. In our fiducial model, this occurs at $T_{\rm gas} > 50$ K, but the exact value depends on $\xi_{\rm H_2}/n_{\rm H}$.
In those regions with such warm gas temperatures, the dust temperature would also be warm.
At $T_{\rm dust} \gtrsim 20$ K, the H$_2$ formation rate may drop considerably, depending on characteristics of chemisorption sites 
on grain surfaces \citep[e.g.,][]{hollenbach71,cazaux04,iqbal14}, which is not considered here.

\section{Scaling relation of the $\hdo$ ratio in the surface ice layers with the atomic D/H ratio}
\label{appendix:water_dh}
Here we derive the scaling relation of the $\hdo$ ratio in the chemically active surface ice layers with the atomic D/H ratio,
which is applicable under the UV irradiation conditions where photodesorption regulates the growth of ice mantles.
Figure \ref{fig:dh_network} shows the important reactions for H$_2$O ice and HDO ice in our models, 
and they are considered in the following analysis.
Let us consider the following set of differential equations which describe temporal variations of the abundances of 
H$_2$O, HDO, OH and OD in the active surface ice layers:
\begin{align}
\ddt \pop{H_2O} =& k_{\rm OH+H_2}\pop{OH}\pop{H_{2}} + k_{\rm OH+H}\pop{OH}\pop{H} - k_{\rm ph}\pop{H_2O} \nonumber \\
                 &- k_{\rm phdes}\pop{H_2O}, \label{eq:apb1} \\
\ddt \pop{HDO} =& k_{\rm OD+H_2}\pop{OD}\pop{H_2} + k_{\rm OD+H}\pop{OD}\pop{H} + k_{\rm OH+D}\pop{OH}\pop{D} \nonumber \\
                &- k_{\rm ph}\pop{HDO} - k_{\rm phdes}\pop{HDO}, \\
\ddt \pop{OH} =& k_{\rm ph}\pop{H_2O} + b_{\rm OH}k_{\rm ph}\pop{HDO} - k_{\rm OH+H_2}\pop{OH}\pop{H_2} \nonumber \\
               &+ k_{\rm O+H}\pop{O}\pop{H} - k_{\rm OH+H}\pop{OH}\pop{H} - k_{\rm OH+D}\pop{OH}\pop{D}, \\
\ddt \pop{OD} =& (1-b_{\rm OH})k_{\rm ph}\pop{HDO} - k_{\rm OD+H_2}\pop{OD}\pop{H_2} \nonumber \\
               & + k_{\rm O+D}\pop{O}\pop{D} - k_{\rm OD+H}\pop{OD}\pop{H},\label{eq:apb4}
\end{align}
where $k_{\rm ph}$ and $k_{\rm phdes}$ are the rate coefficients of photodissociation and photodesorption 
in the active surface ice layers, respectively, of H$_2$O ice and HDO ice.
$b_{\rm OH}$ is the branching ratio to form OH + D for HDO ice photodissociation \citep[$\sim$0.3,][]{koning13}.
%Although species can be buried in the inactive inert ice mantle via the accretion of gaseous species onto the surface,
%we neglect that in Equations (\ref{eq:apb1})-(\ref{eq:apb4}).
%This assumption is verified as seen below.
Combining Equations (\ref{eq:apb1})-(\ref{eq:apb4}), the evolutionary equation of the $\hdo$ ratio in the active ice layers 
($f_{\rm HDO,\,s}=\pop{HDO}/\pop{H_2O}$) can be written 
as follows:
\begin{equation}
\begin{split}
\ddt f_{\rm HDO,\,s} \approx & \frac{d}{dt}\left(\frac{\pop{HDO} + \pop{OD}}{\pop{H_2O} + \pop{OH}}\right)\\
                     \approx & \frac{1}{\pop{H_2O}}\frac{d}{dt}\left(\pop{HDO} + \pop{OD}\right) \\ 
                             & - \frac{\pop{HDO}}{\pop{H_2O}^2}\frac{d}{dt}\left(\pop{H_2O} + \pop{OH}\right) \\
%& \approx -b_{\rm OH}k_{\rm ph}f_{\rm HDO,\,s} + \frac{1}{\pop{H_2O}}[k_{\rm O+D}\pop{O}\pop{D} - k_{\rm O+H}\pop{O}\pop{H}f_{\rm HDO,\,\,s} + k_{\rm OH+D}\pop{OH}\pop{D}]\\
%&\approx -b_{\rm OH}k_2f_{\rm HDO} + \frac{k_3\pop{O}\pop{H}}{\pop{H_2O}}\left[\beta f_{\rm D} - f_{\rm HDO}\right] + \frac{k_5\pop{OH}\pop{H}}{\pop{H_2O}}\beta f_{\rm D}\\
                     \approx & -\left( b_{\rm OH}k_{\rm ph} + \frac{k_{\rm O+H}\pop{O}\pop{H}}{\pop{H_2O}} \right)f_{\rm HDO,\,\,s} \\
                             & + \frac{k_{\rm OH+H}\pop{OH}\pop{H} + k_{\rm O+H}\pop{O}\pop{H}}{\pop{H_2O}}\delta_{\rm hop} f_{\rm D,\,s}, \label{eq:dfhdodt}
\end{split}
\end{equation}
where we used the inequalities $\pop{OH} \ll \pop{H_2O}$, $\pop{OD} \ll \pop{HDO}$, and $f_{\rm HDO,\,s} \ll 1$, which are verified by our numerical simulations.
We defined $f_{\rm D,\,s} = \pop{D}/\pop{H}$.
We also used the relations, $\delta_{\rm hop} \equiv k_{\rm hop}({\rm D})/k_{\rm hop}({\rm H}) \sim k_{\rm O+D}/k_{\rm O+H} \sim k_{\rm OH+D}/k_{\rm OH+H}$,
which should be valid because of the much higher hopping rates of atomic H and D compared to those of atomic O and OH radical. 
The energy barrier difference against hopping between atomic D and H is $\sim$10 K on amorphous water ice \citep{hama12}.
%The time evolution of $f_{\rm D,\,s}$ is controlled by freeze-out of CO and the H$_2$ opr, the timescales of which
%are much longer than the timescale of photoreaction in the regions with low extinction.
%Then, we can consider Equation (\ref{eq:dfhdodt}) as a differential equation of $f_{\rm HDO,\,s}$, where $f_{\rm D,\,s}$ is constant.
When the right hand side of Equation (\ref{eq:dfhdodt}) is less (more) than zero at given $f_{\rm D,\,s}$, 
the ratio $f_{\rm HDO,\,s}$ decreases (increases) on a timescale shorter than $b_{\rm OH}k_{\rm ph}$.
Since the photodissociation timescale in gas with low extinction is much shorter than the dynamical timescale 
($>$1 Myr in the case of our cloud formation model),
let us assume steady state, which corresponds to the minimum (or maximum) of $f_{\rm HDO,\,s}/f_{\rm D,\,s}$.
Then we obtain 
\begin{align}
\frac{f_{\rm HDO,\,s}}{f_{\rm D,\,s}} \approx \delta_{\rm hop} \left( \frac{k_{\rm OH+H}\pop{OH}\pop{H} + 
k_{\rm O+H}\pop{O}\pop{H}}{b_{\rm OH} k_{\rm ph}\pop{H_2O} + k_{\rm O+H}\pop{O}\pop{H}}\right). \label{eq:fhdo_fd}
\end{align}

%Let us evaluate the rhs of Equation (\ref{eq:fhdo_fd}) in the two cases:
%main formation path of water is (i) OH + H$_2$, and (ii) OH + H.
%Unless the reaction of OH + H$_2$ is the rate-determining step of the water formation, 
%$\rate{LH}{OH}$ should be evaluated by formation efficiency of water $(\rate{form}{H_2O})$.
To produce water ice mantles under UV irradiation, the rate of OH formation 
through the hydrogenation of atomic oxygen (i.e., $k_{\rm O+H}\pop{O}\pop{H}$)
should be larger than the photodesorption rate of H$_2$O (see Figure \ref{fig:dh_network}).
When the dynamical timescale is longer than the timescale of the OH formation, the chemistry evolves as 
the OH formation rate and the H$_2$O photodesorption rate are almost balanced:
\begin{align}
k_{\rm O+H}\pop{O}\pop{H} \approx k_{\rm phdes}\pop{H_2O}, \label{eq:rform_water}
\end{align}
This relation is confirmed in our simulations.
%Substituting Equation (\ref{eq:rform_water}) into Equation (\ref{eq:fhdo_fd}), we get
%\begin{align} 
%\frac{f_{\rm HDO,\,s}}{f_{\rm D,\,s}} \approx \frac{\delta_{\rm hop}}{b_{\rm OH}}
%\left( \frac{k_{\rm OH+H}\pop{OH}\pop{H}}{k_{\rm ph}\pop{H_2O}} + P_{\rm phdes}\right). \label{eq:fhdo_fd2}
%\end{align}
The population of OH can be evaluated from the following equation:
\begin{align}
k_{\rm OH+H_2}\pop{OH}\pop{H_2} + k_{\rm OH+H}\pop{OH}\pop{H} = k_{\rm ph}\pop{H_2O} + k_{\rm O+H}\pop{O}\pop{H}. 
\end{align}
Then
\begin{align}
\pop{OH} \approx \frac{k_{\rm ph}\pop{H_2O}}{k_{\rm OH+H}\pop{H} + k_{\rm OH+H_2}\pop{H_2}},\label{eq:pop_oh}
\end{align}
where we used $k_{\rm ph}\pop{H_2O} \gg k_{\rm phdes}\pop{H_2O} \approx k_{\rm O+H}\pop{O}\pop{H}$.
From Equations (\ref{eq:fhdo_fd}), (\ref{eq:rform_water}), and (\ref{eq:pop_oh}), we get the scaling relation,
\begin{align}
f_{\rm HDO,\,s} \approx \frac{\delta_{\rm hop}}{b_{\rm OH}} \left[\frac{1}{1+\Gamma} + P_{\rm phdes} \right] f_{\rm D,\,s}, \label{eq:fhdo_fd_final}
\end{align}
where $P_{\rm phdes}$ is $k_{\rm phdes}/k_{\rm ph} \sim 0.02$ \citep{andersson06,arasa15}.
The term $\delta_{\rm hop} f_{\rm D,\,s}/(1+\Gamma)$ corresponds to $\phdo$ which is discussed in the main text.
In the case with $\Gamma \ll 1$ (or, in other words, OH + H is the dominant formation pathway of H$_2$O ice), we get $f_{\rm HDO,\,s} \approx f_{\rm D,\,s}$.
In another extreme case $\Gamma \gg 1$, we get $f_{\rm HDO,\,s} \approx (\Gamma^{-1} + 0.02)f_{\rm D,\,s}$, 
i.e., $f_{\rm HDO,\,s}$ is much smaller than $f_{\rm D,\,s}$.
For example, $\Gamma$ is $\sim$10$^4$ at $A_V = 1$ mag in our fiducial simulation, and it further increases with the increase of $A_V$.
The minimum of the $f_{\rm HDO,\,s}/f_{\rm D,\,s}$ ratio is $\sim$0.02 in our fiducial simulation, which is well reproduced by Equation (\ref{eq:fhdo_fd_final}).

\section{Comparisons with observations of sulfur-bearing species}
\label{sec:sulfur}
As discussed in Section \ref{sec:metal}, the evolution of the $\op{H_2}$ depends on the assumed heavy metal abundances.
The goal of this appendix is to determine which case, the LM abundances (fiducial case) 
or the HM abundances, gives the better predictions on the $\op{H_2}$ evolution in the ISM.

There are some estimates of $\op{H_2}$ in cold dense clouds/cores from observations of molecules other than H$_2$ in the literature.
We do not use them for the constraint here, because how to estimate $\op{H_2}$ from the observations is not well-established, 
and because the evolution of $\op{H_2}$ in the ISM can be in the non-equilibrium manner and may vary among sources depending on their past physical evolution.
Instead, we compare observations of sulfur-bearing molecules toward diffuse/translucent/molecular clouds in the literature with our model results.
We focus on the total abundance of selected sulfur-bearing molecules in the gas phase (CS, SO, and H$_2$S) and the HCS$^+$/CS abundance ratio.
The former can probe the total abundance of elemental sulfur in the gas phase, 
though the main form of gaseous sulfur is likely to be the neutral atom S or S$^+$ especially in gas with low extinction.
The latter, the HCS$^+$/CS ratio, can probe the ionization degree of the gas, which is related to the S$^+$ abundance (and abundances of the other heavy metal ions).
The HCS$^+$/CS ratio is anticorrelated with the electron abundance in our models, 
because CS is formed by dissociative electron recombination of HCS$^+$ and destroyed by photodissociation.
Although the HCS$^+$/CS ratio also depends on the photodissociation timescale of CS, it is common between the fiducial model and model HM at given $A_V$.
The rate coefficient and the branching ratios for the electron recombination of HCS$^+$ are taken from \citet{montaigne05}.

\citet{turner96} derived the molecular abundances of H$_2$S, CS, and SO in translucent clouds with line of sight visual extinction of up to 5 mag.
He found that the total abundance of these three species with respect to hydrogen nuclei is typically 10$^{-8}$-10$^{-7}$.
In the dense molecular clouds, TMC-1 and L134N, with higher line of sight visual extinction than the samples in \citet{turner96}, 
the total abundance is 10$^{-9}$-10$^{-8}$ \citep{ohishi92,dickens00}.
The lower total abundance in the dense molecular clouds implies that depletion of gaseous sulfur is significant in the dense clouds \citep{joseph86}.
It is not obvious which $A_V$ in our model can be compared with the observations towards the translucent and dense molecular clouds.
Towards TMC-1(CP) and L134N, there is evidence of CO freeze-out; 
the gaseous CO abundance with respect to hydrogen nuclei ($4\times10^{-5}$) is less than the canonical value of 10$^{-4}$ and
infrared absorption by CO ice is detected in the line of sight towards the vicinity of them \citep{whittet07,whittet13}.
We assume that our results at $A_V > 2$ mag, where the gaseous CO abundance decreases to less than $\sim4\times10^{-5}$ because of the freeze-out, 
can be compared with the observations in TMC-1 and L134N, while the results at $A_V < 2$ mag are compared with 
the observations in clouds with lower line of site extinction.

Figure \ref{fig:metal} shows the total abundance of H$_2$S, CS, and SO as a function of visual extinction in our models.
Model HM better reproduces the total abundance of the S-bearing molecules observed in the translucent clouds, 
while the fiducial model better reproduces the total abundance observed in the dense molecular clouds.
This result implies that again, the depletion of sulfur from the gas phase occurs during the formation and evolution of molecular clouds, 
and that model HM underestimates the degree of the sulfur depletion.

The HCS$^+$/CS ratio has been derived toward diffuse clouds \citep[0.08,][]{lucas02}, 
clouds with line of sight visual extinction of up to 5 mag \citep[0.01-0.1,][]{turner96}, 
and the dense molecular clouds, TMC-1 and L134N \citep[0.06,][]{ohishi92}.
Roughly speaking, the ratio is in the range of 0.01-0.1 in the diffuse and dense ISM.
Figure \ref{fig:metal2} shows the HCS$^+$/CS ratio and the electron abundance in the fiducial model and model HM.
The fiducial model reproduces the HCS$^+$/CS ratio much better than model HM, 
though the both models tend to underestimate the HCS$^+$/CS ratio compared with observations, i.e., overestimate the ionization degree of the gas.

Considering the HCS$^+$/CS ratio is reproduced by the fiducial model much better than by model HM,
we conclude that the fiducial model gives a better prediction for the S-bearing species and thus for the $\op{H_2}$.
In particular, model HM seems to overpredict the S-bearing species in the gas phase.
The non-success of the model HM probably means that a non-negligible fraction of sulfur is incorporated into dust grains \citep{keller02}
and/or the non-thermal desorption rates of icy sulfur are overestimated in the current model; 
the partitioning of elemental sulfur between gas and ice in our model depends on the assumed non-thermal desorption rates, 
especially chemisorption probabilities, which remain uncertain at the current stage.
In our models, the dominant sulfur reservoirs in ices are HS and H$_2$S, though there has been no detection of H$_2$S ice in the ISM.
HS ice is formed from H$_2$S ice via the hydrogen abstraction reaction, H$_2$S + H, with an activation energy barrier of 860 K \citep{hasegawa92}.
HS ice is hydrogenated to form H$_2$S ice again, desorbing $\sim$1 \% of the product H$_2$S through chemisorption.
This loop is the main mechanism of the desorption of icy sulfur in our model as in \citet{garrod07}.
If we set the chemisorption probability for the hydrogenation of HS ice to be 0.1 \%, only 3 \% of elemental sulfur remains 
in the gas phase at $A_V = 3$ mag in model HM, though gaseous sulfur is still more abundant than in the fiducial model.

\end{appendix}

\end{document}